\documentclass[12pt]{article}
\usepackage{amsfonts,amsmath,amssymb}   
\usepackage[english]{babel}             
\usepackage{lscape}                     
\usepackage[vcentermath]{youngtab}      
\usepackage{longtable}                  
\usepackage{pstricks,pst-node,satake}   
\usepackage{cite}

\newcommand{\cx}{\mathbb{C}}
\newcommand{\N}{\mathcal{N}}
\newcommand{\M}{\mathcal{M}}
\newcommand{\fg}{\mathfrak{g}}
\newcommand{\ff}{\mathfrak{f}}

\newcommand{\fe}{\mathfrak{e}}
\newcommand{\fsu}{\mathfrak{su}}
\newcommand{\fso}{\mathfrak{so}}
\newcommand{\fsl}{\mathfrak{sl}}

\newcommand{\fund}{\tiny \yng(1)}
\newcommand{\twoform}{\tiny \yng(1,1)}
\newcommand{\threeform}{\tiny \yng(1,1,1)}
\newcommand{\fourform}{\tiny \yng(1,1,1,1)}

\newcommand{\symm}{\tiny \yng(2)}

\newcommand{\threehook}{\tiny \yng(2,1,1)}

\newcommand{\fivehook}{\tiny \yng(2,1,1,1,1)}

\newcommand{\eps}{\epsilon}
\newcommand{\beps}{\bar{\epsilon}}

\begin{document}

\begin{titlepage}

\begin{flushright}
  ULB-TH/08-11 \\
  UB-ECM-PF-08-10
\end{flushright}

\begin{center}
\vspace{3cm} \baselineskip=16pt {\LARGE \bf Extended Symmetries in
Supergravity: \\ \vspace{.3cm} the Semi-simple Case}
 \vskip 1 cm
{\large  Axel Kleinschmidt$^1$ and Diederik Roest$^2$} \\
\vskip 10 mm {\small
$^1$ Physique Th\'eorique et Math\'ematique,\\
Universit\'e Libre de Bruxelles \&{} International Solvay Institutes,\\
Boulevard du Triomphe, ULB -- CP 231, B-1050 Bruxelles, Belgium\\[3mm]
   $^2$ Departament Estructura i Constituents de la Materia \\
    \& Institut de Ci\`{e}ncies del Cosmos, \\
    Universitat de Barcelona, \\
    Diagonal 647, 08028 Barcelona, Spain \\ \vspace{.2cm}
   E-mails: {\tt axel.kleinschmidt@ulb.ac.be}, {\tt droest@ecm.ub.es}}
\end{center}

\bigskip
\centerline{ABSTRACT}
\bigskip\bigskip

The bosonic sector of various supergravity theories reduces to a
homogeneous space $G/H$ in three dimensions. The corresponding
algebras $\fg$ are simple for (half-)maximal supergravity, but can
be semi-simple for other theories. We extend the existing literature
on the Kac--Moody extensions of simple Lie algebras to the
semi-simple case. Furthermore, we argue that for $\N = 2$
supergravity the simple algebras have to be augmented with an
$\fsu(2)$ factor.

\end{titlepage}

\section{Introduction}

One of the intriguing aspects of supergravities are their
hidden symmetries. Upon dimensional reduction over a torus $T^d$ of
any theory containing gravity, one would expect an $SL(d)$ global
symmetry in four or more dimensions and $SL(d+1)$ in three
dimensions. These are lower-dimensional remnants of the
diffeomorphisms on the internal torus. In addition there will be
generators that originate from e.g.~higher-dimensional gauge
symmetries. The surprising feature of supergravity is that these two
types of symmetries combine to form a much larger symmetry group.
For instance, for maximal supergravity the global symmetries are the
exceptional groups\footnote{The groups and algebras of this paper
are of split real form unless explicit compact notation is used.}
$E_{11-D}$ in $D\ge 3$ dimensions \cite{Cremmer-Julia, Julia81}.

The phenomenon of hidden symmetries is perhaps most striking in
three dimensions. All propagating degrees of freedom can be
described by scalars and, in the cases of interest, these transform
in a non-linear representation under a global symmetry. That is,
the bosonic sector of the theory consists of gravity (which is
non-propagating in three dimensions) coupled to a scalar coset $G / H$,
where $G$ is the global symmetry group and $H$ is its maximal
compact subgroup. For instance, maximal supergravity in three
dimensions reduces to the coset
 \begin{align}
  \frac GH = \frac{E_{8}}{SO(16)} \,, \label{G/H-max}
 \end{align}
while half-maximal supergravity is given by
 \begin{align}
  \frac GH = \frac{SO(8,8+n_{\rm V})}{SO(8) \times SO(8+n_{\rm V})} \,, \label{G/H-halfmax}
 \end{align}
where $n_V$ corresponds to the number of vector multiplets in ten
dimensions.

So far our discussion has been concerned with {\em finite-dimensional}
hidden symmetries, generated by a {\em simple} Lie algebra $\fg$.
Yet more intriguing are the results and conjectures on extended symmetries,
featuring the infinite-dimensional Kac--Moody extensions of the simple
$\fg$. For instance, it has been 
proven that the affine extension $\fg^+$ appears upon reduction to
two dimensions \cite{Julia82, Geroch, Breitenlohner:1986um,Nicolai:1987kz}. In
particular, all bosonic solutions in $D=2$ form a non-linear
representation of $\fg^+$. The over- and very-extensions $\fg^{++}$
and $\fg^{+++}$ have been conjectured to play a role in $D=1$ and
$D=0$, respectively \cite{Julia82, Nicolai}.
Finally, further conjectures have been made about the role of the
latter two Kac--Moody algebras in relation to the full supergravity
without dimensional reduction to $D <2$, see
\cite{damourhenneauxnicolai, damournicolai, tenforms1} and
\cite{West1, West2, West3} respectively.

We will not be concerned with the dynamical realisation of the
Kac--Moody symmetries according to these different proposals.
Instead, we will focus on a necessary requirement for these
conjectures to work: the algebraic correspondence between
bosonic supergravity fields and a well-defined truncation of the Kac--Moody
generators, and we focus on the very-extended algebras.\footnote{Our construction of extended algebras in the semi-simple case and the analysis of their spectrum is also valid for the over-extended case.}
This matching has been performed for the physical
degrees of freedom of many supergravities, see
e.g.~\cite{West:2002jj,Nicolai:2003fw,Kleinschmidt, Riccioni:2008}.

This correspondence can be extended to
non-propagating supergravity fields. More concretely, many
supergravities allow for the introduction of certain $(D-1)$- and $D$-form
potentials. The former can be seen as the duals of deformation
parameters that introduce massive deformations or gaugings to the
theory. The latter can correspond to certain constraints that have to be
imposed on the gauge parameters for consistency of the gaugings.
These non-propagating degrees of freedom will be referred to as deformation potentials and top-form potentials, respectively.
Interestingly, it was found recently that the Kac--Moody
algebra $\fg^{+++}$ contains exactly the right generators to
correspond to the deformation and top-form potentials of
maximal and half-maximal supergravity in all dimensions \cite{gaugings1, gaugings2,
half-maximal} (see also~\cite{tenforms1, IIB-10forms, West-tenforms} for earlier results in ten dimensions and \cite{Riccioni:2007ni,Bergshoeff:2008qd} for a detailed analysis of the algebraic structure in the gauged case).

In this paper we want to address the natural and interesting question
to what extent the above results hold for other supergravities as
well. We will mainly be concerned with supergravity theories that
can be formulated in four or more dimensions\footnote{For this
reason we will adhere to four-dimensional notation for $\N$,
i.e.~the number of supercharges of a theory is $4 \N$ in any
dimension. }, and not those that only live in three dimensions.
Nevertheless, as stressed before, it will be crucial for our
analysis to understand what the various supergravities reduce to in
three dimensions. The bosonic sector of any supergravity reduces to
a scalar manifold coupled to gravity, and different amounts of
supersymmetry impose different constraints on this scalar manifold
\cite{Tollsten}. In particular, the bosonic sectors of all
supergravity theories with more than eight supercharges, i.e.~with
$\N > 2$, reduce to a homogeneous space $G/H$ in three dimensions.
Examples are the maximal and half-maximal supergravities given in
\eqref{G/H-max} and \eqref{G/H-halfmax}, but this result also holds
for the `exceptional' supergravity theories with $10, 12, 18, 20$ or
$24$ supercharges~\cite{Gunaydin:1983rk}.

For theories with $\N \leq 2$, i.e.~with eight or less real supercharges,
one encounters more general scalar manifolds than homogeneous ones.
Nevertheless, the subset of theories for which the scalar manifold
is homogeneous is more tractable and still interesting, and has
proven very valuable in many applications. In particular, one can
study the hidden symmetries and ask similar questions about the
corresponding Kac--Moody extensions as discussed above in the context
of maximal and half-maximal supergravity. In this paper we will only
be concerned with the theories at these `points of homogeneity' in
the moduli space of $\N \leq 2$ theories.

Two points are important to notice in the context of this paper. The
first concerns the algebra $\fg$ that is generated by the group $G$ of isometries
of the homogeneous spaces in three dimensions. As can be read off
from \eqref{G/H-max} and \eqref{G/H-halfmax}, this is always a
simple algebra for maximal and half-maximal supergravity. In fact
this holds for all supergravity theories with $\N >2$. A simple Lie
algebra can straightforwardly be promoted to a Kac--Moody algebra by
affine, over- and very-extensions
\cite{Goddard:1983at,Gaberdiel:2002db}. In contrast, for $\N \leq 2$
the algebra of isometries $\fg$ is not necessarily simple but can be
semi-simple as well.\footnote{There are also (non-symmetric) homogeneous spaces
with non-semi-simple groups of isometries, see e.g.~\cite{deWit}, but we
will not consider these here.} In fact, as we will see in sections
\ref{sec:sugra} and \ref{sec:discussion}, it can be argued that it
is generically semi-simple for homogeneous scalar manifolds in  $\N
= 2$ and $\N = 1$ supergravities. One thus needs a proposal for the
corresponding affine, over- and very-extensions in the semi-simple
case. This will be provided in section \ref{sec:extensions}.

\begin{table}[ht]
\begin{center}
\begin{tabular}{||c||c|c||}
\hline $D$ & $H_{\rm R}$ & with \\ \hline \hline
 $11$ &  $1$ & \\ \hline
 $10$ & $SO(n_1) \times SO(n_2)$ & $ n_1+n_2 = \N/4$ \\ \hline
 $9$ & $O(n)$ & $n = \N /4$ \\ \hline
 $8$ & $U(n)$ & $n = \N /4$ \\ \hline
 $7$ & $Sp(n)$ & $n = \N /4$ \\ \hline
 $6$ & $Sp(n_1) \times Sp(n_2)$ & $ n_1+n_2 = \N/2 \,\, (n_{1,2} \leq 2)$ \\ \hline
 $5$ & $Sp(n)$ & $n = \N /2$ \\ \hline
 $4$ & $U(n)$ & $n = \N $ \\ \hline
 $3$ & $SO(n)$ & $n = 2 \N $ \\
\hline
\end{tabular}
\caption{\sl The R-symmetry groups of supergravities for various values
of $D$ and $\N$, adapted from \cite{Strathdee,Keurentjes:2002xc}, and where
$Sp(n)$ denotes the compact symplectic group of dimension $n (2n+1)$
(sometimes also denoted in the literature as $USp(2n)$). In ten dimensions
there are two possibilities for 
maximal supergravity while in six dimensions there are two
possibilities for half-maximal supergravity. In both cases the
non-chiral theory has $n_1=n_2$ while the chiral one has $n_1 n_2 =
0$.} \label{tab:R}
\end{center}
\end{table}

The second point concerns the R-symmetry group, i.e.~the global
symmetry that rotates the different supercharges of a supersymmetric
theory. The R-symmetry groups for different values of $D$ and $\N$
have been summarised in table \ref{tab:R}. It turns out that the
R-symmetry group $H_{\rm R}$ of maximal supergravity coincides with
the compact part of the global symmetry group, i.e.~we have $H_{\rm
R} = H$. This can be checked for $D=3$ using \eqref{G/H-max} but
holds also in higher dimensions.\footnote{In dimensions lower than $D=3$
  the R-symmetry group $H_{\rm R}$ of maximal supergravity is again identical to $H$,
  though $H$ is now infinite-dimensional  \cite{Nicolai:2004nv,Damour:2005zs,deBuyl:2005mt,Kleinschmidt:2007zd}.}
The same is true for half-maximal supergravity if one only considers
the graviton multiplet (corresponding to $n_{\rm V}=-7$ in
\eqref{G/H-halfmax}). In the presence of additional vector
multiplets (or tensor multiplets in the six-dimensional chiral
theory) the global symmetry group and its compact part are larger.
Again this also holds for the `exceptional' supergravities. Hence
for $\N > 2$ one always has $H_{\rm R} \subseteq  H$.

For $\N =2$, however, this is not always the case. In particular, in
the absence of hyper multiplets there is an $SU(2) \subseteq  H_{\rm
R}$ part missing in $H$ for all dimensions. For example, pure $\N=2$
supergravity in $D=5$ should have $SU(2)$ R-symmetry but there are
no scalars giving a scalar manifold $G/H$ with $SU(2)\subseteq H$. For
precisely such cases there is also a problem with the correspondence
between the Kac--Moody algebra and supergravity, as the former does
not contain all the potential gaugings of the latter. In particular,
the potentials corresponding to the gaugings of $SU(2) \subseteq
H_{\rm R}$ in $D\leq 5$ are not present in the Kac--Moody algebra.
This mismatch has been noted in \cite{Gomis}. In section
\ref{sec:sugra} a resolution is proposed by including an additional
`empty' $SU(2) / SU(2)$ scalar manifold, such that $H_{\rm R}
\subseteq  H$ holds for these cases as well.\footnote{Such a factor appears in \cite{Cremmer80} but apparently has been replaced by "1" in the subsequent literature.}
 Of course, the additional
compact factor does not introduce any physical degrees of freedom.
However, we will see that the corresponding extended semi-simple
algebra does contain the possible gaugings of this compact factor,
and agrees perfectly with the results of \cite{Gomis}.\footnote{Similarly, the
absence of four-forms in $\fg_2^{++}$ led to a paradox
concerning higher-order corrections to this five-dimensional supergravity
\cite{Mizoguchi}. We expect our proposal to resolve this puzzle as well.}

Summarising, the purpose of this paper is twofold:
 \begin{itemize}
 \item Firstly, we make a proposal
for the extensions for semi-simple $\fg$, i.e.~the analogon of the
affine, over- and very-extension of simple $\fg$. The corresponding
extensions will turn out to be quotients of certain derived Kac--Moody
algebras. We will present a number of arguments why these extensions
are the relevant ones in the context of supergravity.
 \item
Secondly, we argue that the problematic case with $H_{\rm R} \not
\subseteq  H$ can be remedied by the extension of the scalar coset
with the missing compact factor. For $\N = 2$ supergravity without
hyper multiplets this is an $SU(2) / SU(2)$ factor. Also in the
absence of hyper multiplets one then has $H_{\rm R} \subseteq H$ and
$\fg$ semi-simple in three dimensions. (In the presence of hyper
multiplets $H_{\rm R} \subseteq H$ follows directly from the scalar
manifold of the hyper multiplets.)
 \end{itemize}

The outline of this paper is as follows. In section
\ref{sec:extensions} we will present a proposal for the extensions
of semi-simple Lie algebras. In section \ref{sec:sugra} these will
be applied to a pair of $\N=2$ supergravity examples: one with and
one without hyper multiplets. In section \ref{sec:E11} the relation
of these examples to $\fe_{11}$ is discussed. Our conclusions are presented
in section \ref{sec:discussion}. In appendix \ref{sec:five-forms} we
review the supersymmetry algebra of pure $\N=2$ supergravity in
$D=5$ and show the possibility to include certain five-forms. Some general remarks and a particular example of $\N=1$ supergravity are discussed in appendix~\ref{sec:N=1}.
Finally, appendix \ref{sec:tables} contains the decomposition tables
of the Kac--Moody algebras corresponding to the supergravity
examples.

\section{Extensions of semi-simple Lie algebras} \label{sec:extensions}

In this section we discuss the general problem of obtaining Kac--Moody
extensions of direct sums of finite-dimensional simple Lie algebras.

\subsection{Review of the extension process for simple Lie algebras}

For a complex, finite-dimensional and simple Lie algebra $\fg$
there exists a standard process of
extending the Dynkin diagram by three nodes to obtain the so-called
very-extension $\fg^{+++}$ \cite{Goddard:1983at,Gaberdiel:2002db}. This
extension process consists of three steps,
where the first additional node leads to the so-called non-twisted affine
extension which we denote here as $\fg^+$. The way the affine node is attached
to the Dynkin diagram of $\fg$ is governed by the highest root of $\fg$. A
list of the diagrams of all non-twisted $\fg^+$ can be found for example in \cite{Kac}. As a
vector space $\fg^+$ is
isomorphic to $\fg[[t,t^{-1}]]\oplus \cx c \oplus \cx d$, i.e. the centrally
extended loop algebra over $\fg$ with spectral parameter $t$, central element
$c$ and derivation $-t\frac{d}{dt}$. From the
algebraic point of view the derivation $d$ serves to desingularize the inner
product $(\cdot|\cdot)$ on the Cartan subalgebra. Since $c$ is central it
satisfies $(c|h)=0$ for all Cartan generators $h$ of the finite-dimensional
$\fg$ and also $(c|c)=0$. By introducing the derivation $d$ with  $(d|c) = -1$
this degeneracy of the inner product is alleviated.\footnote{The elements $c$
  and $d$ can be thought of as two independent light-cone coordinates.}
If $\fg$ is of rank $r$, so that there are $r$ independent Cartan generators
in $\fg$, the affine extension $\fg^{+}$ has $r+2$ commuting
diagonal elements, exceeding the number of nodes of the Dynkin diagram by
one. We will always use the notation $\fg^+$ in this paper to refer to the
Kac--Moody algebra $\fg[[t,t^{-1}]]\oplus \cx c \oplus \cx d$ and call it the
affine version of the simple Lie algebra $\fg$.

The next step in the extension process leads to the over-extension $\fg^{++}$
and can be thought of as giving a Dynkin-diagrammatic home to the derivation
$d$ so that the number of nodes agrees again with the number of independent
diagonal elements. In order to ensure all the properties of $d$ on the affine
subalgebra $\fg^+$ the new node has to be joined with a single undirected line
to the affine node. The resulting algebra has inner product of Lorentzian
signature on the Cartan subalgebra.\footnote{In many physically interesting cases the
associated Kac--Moody algebra is hyperbolic such that the BKL limit near a
space-like singularity will exhibit chaos
\cite{Damour:2000wm,Damour:2001sa,Damour:2002fz}.}
In the last step (very-extension) one adjoins another node with a single line to the hyperbolic
node to obtain the Lorentzian algebra $\fg^{+++}$
\cite{Gaberdiel:2002db}. As the number of Cartan subalgebra elements will be
of importance for our proposal for extending semi-simple algebras we summarize
this again: For simple $\fg$ of rank $r$ there are $r+2$, $r+2$ and $r+3$
independent Cartan subalgebra generators for $\fg^+$, $\fg^{++}$ and
$\fg^{+++}$, respectively. In contrast, the loop algebra $\fg[[t,t^{-1}]]$ has
$r$ Cartan subalgebra elements.

The affine algebra $\fg^+$ arises from gravity models for simple $\fg$ as
follows. We consider gravity in $D=3$ coupled to a $G/H$ scalar
coset, where we now assume that a real form of $\fg$ has been chosen which is
the Lie algebra of $G$. As always, $H=K(G)$ is the maximal compact subgroup of
$G$ and describes the local symmetries. In the reduction to $D=2$ the affine
extension $\fg^+$ arises as the new and larger symmetry algebra as can be
shown by considering a linear system based on $\fg^+$
\cite{Breitenlohner:1986um,Nicolai:1991tt}. This symmetry is
commonly referred to as Geroch symmetry \cite{Geroch:1970nt}; the
importance of the central extension and derivation were first noticed in
\cite{Julia:1981wc,Julia:1982fs}. The central element is
related to the size of the non-compact two-dimensional space-time, whereas the
derivation $d$ is related to the size of the circle used in the reduction from
$D=3$ to $D=2$.\footnote{More precisely, in conformal gauge the $D=2$ metric
  has as its one independent component the conformal factor which is acted upon
  by symmetry transformations in the $c$ direction. If the $G/H$ coset arises
  from the reduction of some higher dimensional model $d$ acts on the overall
  size of all compact direction.} The spectral parameter $t$ is needed to
distinguish and organise the infinity of independent auxiliary
scalar fields dual to the reduced scalars one can introduce in $D=2$
(the so-called dual potentials \cite{Breitenlohner:1986um,Nicolai:1991tt}).

\subsection{Extending semi-simple Lie algebras}

If one instead starts with a direct sum, say $\fg_a \oplus \fg_b$ of simple
finite-dimensional Lie algebras rather than a single simple Lie
algebra, the extension process described above is not uniquely
defined any longer. The ambiguity arises already for the affine
extension when defining $(\fg_a\oplus\fg_b)^+$. One possibility
would be to define $(\fg_a\oplus\fg_b)^+ = \fg_a^+\oplus \fg_b^+$,
treating the two algebras also completely independently in the
extension process. This results in two independent central elements
and two independent derivations for the two summands. However, the
loop algebra construction also suggests another possibility, namely
to consider $(\fg_a\oplus\fg_b)^+ = (\fg_a\oplus
\fg_b)[[t,t^{-1}]]\oplus \cx c \oplus \cx d$. In this case there is
a {\em common} spectral parameter, a single common derivation $d=-t\frac{d}{dt}$
and a common central element. We note that this definition
does not give a Kac--Moody algebra; it is closely related to
the Kac--Moody algebra $\fg_a^+\oplus\fg_b^+$ from which it differs by having
the two central charges $c_a$ and $c_b$ as well as the two derivations $d_a$
and $d_b$ identified.

{}From the perspective of the connection to (super-)gravity theories the
second (non Kac--Moody) option is preferred when one repeats the arguments
leading to the emergence of the affine symmetry reviewed above. Even if
$\fg$ is not simple the reduction to $D=2$ should still give rise to only one
central element and one derivation since they have a geometric
origin. Therefore the symmetry consideration of gravity coupled to
scalar cosets leads to the second option of defining the affine
extension of a semi-simple $\fg_a\oplus \fg_b$.

\begin{figure}
\centering
\begin{picture}(410,160)
\thicklines
\put(-10,10){$(a)$ Finite}
\put(10,60){\circle{40}}\put(6,58){$\fg_a$}
\put(10,120){\circle{40}}\put(6,118){$\fg_b$}
\put(75,10){$(b)$ Affine}
\put(100,60){\oval(60,40)}\put(95,58){$\fg_a^+$} \put(130,60){\circle*{10}}
\put(100,120){\oval(60,40)} \put(95,118){$\fg_b^{+}$}\put(130,120){\circle*{10}}
\put(175,10){$(c)$ Over}
\put(195,60){\oval(60,40)}\put(190,58){$\fg_a^+$} \put(225,60){\circle*{10}}
\put(195,120){\oval(60,40)} \put(190,118){$\fg_b^{+}$}\put(225,120){\circle*{10}}
\put(255,90){\circle*{10}}\put(225,60){\line(1,1){30}}\put(225,120){\line(1,-1){30}}
\put(310,10){$(d)$ Very}
\put(320,60){\oval(60,40)}\put(315,58){$\fg_a^+$} \put(350,60){\circle*{10}}
\put(320,120){\oval(60,40)} \put(315,118){$\fg_b^{+}$}\put(350,120){\circle*{10}}
\put(380,90){\circle*{10}}\put(350,60){\line(1,1){30}}\put(350,120){\line(1,-1){30}}
\put(410,90){\circle*{10}}\put(380,90){\line(1,0){30}}
\multiput(55,0)(0,10){16}{\line(0,1){5}}
\multiput(152,0)(0,10){16}{\line(0,1){5}}
\multiput(275,0)(0,10){16}{\line(0,1){5}}
\end{picture}
\caption{\label{ssppp} \sl The extension process for semi-simple
  $\fg_a\oplus\fg_b$ in terms of Dynkin diagrams. The marked nodes on the
  blobs for the  affine algebras correspond to the affine nodes of the
  non-twisted affine extensions of the simple $\fg_a$ and $\fg_b$.
  As explained in the text the simple algebras
  $(\fg_a\oplus\fg_b)^{++}$ and   $(\fg_a\oplus\fg_b)^{+++}$ are obtained from
  these Dynkin diagrams by taking the quotient of the derived algebra by its
  center. This amounts to removing the derivation and central element present
  in the Kac--Moody algebras described by these Dynkin diagrams. The affine
  algebra $(\fg_a\oplus\fg_b)^{++}$ is defined as in (\ref{affg1g2}). }
\end{figure}
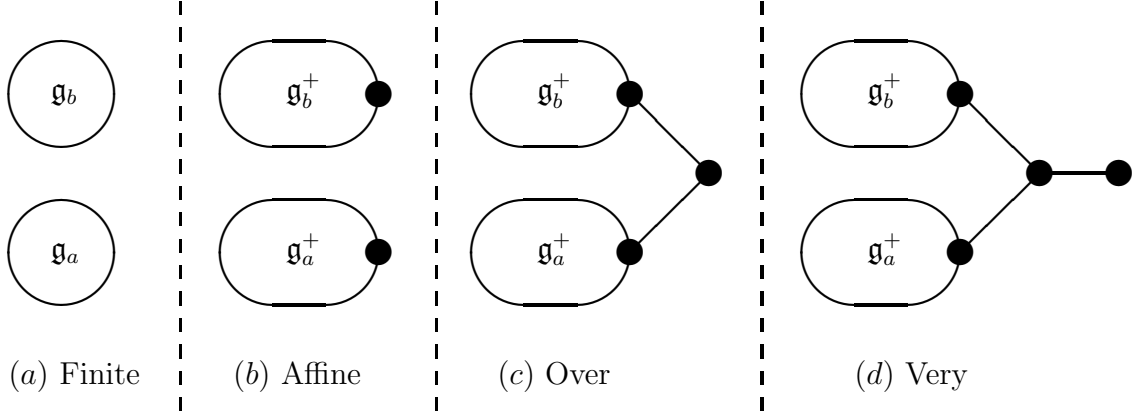

For this reason we will adopt from now on the definition
\begin{equation}\label{affg1g2}
(\fg_a\oplus\fg_b)^+ := (\fg_a\oplus \fg_b)[[t,t^{-1}]]
   \oplus \cx c\oplus \cx d\,.
\end{equation}
Of course, in terms of loop algebras one has
$(\fg_a\oplus \fg_b)[[t,t^{-1}]]= \fg_a[[t,t^{-1}]]\oplus
\fg_b[[t,t^{-1}]]$. As we have stressed, equation (\ref{affg1g2}) does not
correspond to a Kac--Moody algebra but is related to the Kac--Moody algebra
$\fg_a^+\oplus \fg_b^+$, the Dynkin diagram of which is displayed in
fig.~\ref{ssppp}(b), by identifying the two central elements with
each other and by identifying also the two derivations.

Continuing to the over-extension $(\fg_a\oplus\fg_b)^{++}$ it is
natural to construct a Kac--Moody algebra where the single derivation $d$
of (\ref{affg1g2}) is included naturally as for the simple case.
Since the derivation acts
on both loop algebras $\fg_a[[t,t^{-1}]]$ and $\fg_b[[t,t^{-1}]]$
alike, the two disjoint Dynkin diagrams should be joined via the
hyperbolic node, see fig.~\ref{ssppp}(c). The Cartan element of this
common hyperbolic node will act as the derivation on both loop
algebras if the link is a single line and undirected. The
generalised Cartan matrix encoded by the diagram of
fig.~\ref{ssppp}(c) has one zero eigenvalue which corresponds to the
diagonal generator $\tilde{c}=c_a-c_b$ in terms of the central
elements $c_a$ and $c_b$ of the constituent $\fg_a^+$ and $\fg_b^+$.
The Kac--Moody algebra $\tilde{\fg}^{++}$ defined by
fig.~\ref{ssppp}(c) requires therefore a new derivation $\tilde{d}$ to
desingularize the inner product which is degenerate due to the
presence of $\tilde{c}$. Furthermore, the derived algebra
$(\tilde{\fg}^{++})'=[\tilde{\fg}^{++},\tilde{\fg}^{++}]$ of the
Kac--Moody algebra $\tilde{\fg}^{++}$ is isomorphic to
$\tilde{\fg}^{++}$ without $\tilde{d}$.  The generator $\tilde{c}=c_a-c_b$ is
central in $\tilde{\fg}^{++}$. Taking the quotient of the derived algebra by
the center spanned by $\tilde{c}$ will lead to a simple,
infinite-dimensional Lie-algebra \cite{Kac} which we will call the
over-extension $(\fg_a\oplus\fg_b)^{++}$ of the semi-simple algebra
$\fg_a\oplus\fg_b$:
\begin{equation}
(\fg_a\oplus\fg_b)^{++} := (\tilde\fg^{++})'/\langle \tilde{c}\rangle.
\end{equation}
It is {\em not} a Kac--Moody algebra but differs from
one, i.e.~ $\tilde{\fg}^{++}$, only in the diagonal elements in the same
way that a loop algebra $\fg[[t,t^{-1}]]$ differs from the affine
algebra $\fg^+$. Moreover, the quotient by the one-dimensional space spanned
by  $\tilde{c}=c_a-c_b$ identifies the two individual central elements of the
affine algebras $\fg_a^+$ and $\fg_b^+$ which is exactly what we argued for on 
the basis of the Geroch symmetry in $D=2$ gravity.

From the construction of $(\fg_a\oplus\fg_b)^{++}$ it is now easy to
construct the very-extension $(\fg_a\oplus\fg_b)^{+++}$ by adding
another node to the Dynkin diagram. It is again understood that
$(\fg_a\oplus\fg_b)^{+++}$ is constructed from the corresponding
Kac--Moody algebra $\tilde{\fg}^{+++}$, with Dynkin diagram shown in
fig.~\ref{ssppp}(d), by removing the derivation and central element
in the same way as described above.\footnote{It can be seen that
$c_a-c_b$ is still central in   $\tilde{\fg}^{+++}$ and that the center is
one-dimensional.} 
By abuse of notation we will sometimes refer to the Dynkin diagram of
$\tilde{\fg}^{+++}$ as the Dynkin diagram of $(\fg_a\oplus \fg_b)^{+++}$. 

There is a natural extension of our proposal to the case $\fg_a
\oplus \fg_b \oplus \fg_c \oplus \ldots$, consisting of $n$ simple
factors. On general grounds there will now be $n-1$ central elements
$\tilde{c}_a = c_a - c_b$, $\tilde{c}_b = c_b - c_c$, $\ldots$ such
that again in $(\fg_a \oplus \fg_b \oplus \fg_c \oplus \ldots)^{++}$
all originally distinct central charges are identified in agreement
with the Geroch symmetry. This extends also to the very-extended case.

The very-extended algebras $(\fg_a\oplus\fg_b)^{+++}$ we construct
in the fashion described above have $r_a+r_b+3$ diagonal elements
for finite-dimensional $\fg_a$ and $\fg_b$ of rank $r_a$ and $r_b$,
respectively. The signature of the inner product on these diagonal
elements is $(r_a+r_b+2,1)$ and therefore of Lorentzian type.\footnote{This
  follows since there is an $(r_a+r_b+2)$-dimensional space-like subspace, a
  one-dimensional kernel and at least a one-dimensional time-like subspace. A
  dimension count then shows that the quotient by the kernel has signature
  $(r_a+r_b+2,1)$.}

We note that one can also use different real forms of $\fg_a$ and
$\fg_b$ in this extension process, they do not need to be in split
real form. However, we demand from the symmetries of the reduction
of $D=3$ gravity that the affine, over- and very-extended nodes be
non-compact. Of course, the resulting diagram then should be an
allowed almost split real form of the very-extended Kac--Moody
algebra, see e.g.~\cite{Back-Valente:1995,BenMessaoud:2006}. The properly
defined Weyl groups of the quotient Lie algebras introduced in this section
should be related to U-duality in low dimensions~\cite{Obers:1998fb}.

\section{Applications to $\N = 2$ supergravity} \label{sec:sugra}

In this section we will discuss the emergence of semi-simple Lie
algebras in $\N = 2$ supergravity, and analyse in detail a pair of
examples of very-extended semi-simple Lie algebras. These illustrate
our proposal of the previous section and in addition will provide a
number of consistency checks.

\subsection{General aspects} \label{sec:N=2general}

As discussed in the introduction, the $\N = 2$ theories have the highest number of
supercharges that allow for inhomogenous scalar manifolds. Of
course $\N = 2$ supersymmetry does impose a number of geometric
restrictions on these spaces. Most importantly, the scalar manifolds
split up in two parts, parametrised by scalars of the
vector\footnote{The vector multiplet does not comprise any scalars
in six dimensions, but instead there is a tensor multiplet which
does. For simplicity we will refer to the corresponding scalar
manifold as $\M_V$ as well.} and hyper multiplets, respectively:
 \begin{align}
  \M = \M_V \times \M_H \,.
  \label{MV-MH}
 \end{align}
Further requirements on the scalar manifold $\M_V$ are
dimension-dependent: it is very special real, special K\"{a}hler and
quaternionic-K\"ahler in $D=5,4,3$, respectively. In $D=6$ it is
given by a particular homogeneous space. The scalar manifold $\M_H$
is quaternionic-K\"ahler in any dimension $3 \leq D \leq 6$. See
e.g.~\cite{deWit} and references therein for further details on these spaces.

We will be concerned with the subset of $\N = 2$ supergravities
whose scalar manifolds are homogeneous spaces. These have been
classified in \cite{homogeneous1, homogeneous2}. In particular we
are interested in the global symmetries of these theories. From the
split of scalar manifolds \eqref{MV-MH} it follows that the
symmetries will generically be semi-simple. Only in the absence of
hyper multiplets\footnote{We will not consider the case of only
hyper multiplets as these theories only live in $D=3$.} can one have
a simple\footnote{The Kac--Moody extensions of simple algebras
associated to $\N =2$ theories without hyper multiplets have been
discussed in \cite{Kleinschmidt, Riccioni:2008}.} $G$.

Furthermore, the R-symmetry is not always contained in the compact
subgroups of $G$ for $\N = 2$ supergravities. This is easiest to see
in three dimensions where one has only vector and hyper multiplets.
Both multiplets consist of the same fields, being four scalars and
two dilatini. They only differ in the way they transform under the
R-symmetry, which is $H_{\rm R} = SO(4) \simeq SU(2) \times SU(2)$:
only one $SU(2)$ factor acts on the scalars of the vector
multiplets, while the other factor only acts on the scalars of the
hyper multiplets \cite{Tollsten}. This implies that both factors of $H_{\rm
  R}$ will
be contained in $H$ if and only if hyper multiplets are present as
well. The same will hold for the possible uplift of these theories to higher
dimensions.

Although the reasoning is completely general and will apply to all
$\N=2$ theories, it may be instructive to consider specific
examples of both kinds (i.e.~with and without hyper multiplets):
\begin{itemize}
\item Pure
$\N=2$ supergravity in five dimensions consists of only the graviton
multiplet. Its bosonic sector comprises the graviton and a vector,
while its fermions are a pair of symplectic Majorana gravitini.
\item One
can couple this theory to seven hyper multiplets whose scalars
parametrise an $F_{4} / (SU(2) \times Sp(3))$ scalar coset.
\end{itemize}
These will be referred to as the pure theory and the coupled theory, respectively.
Interestingly, both can be
obtained as a truncation of $\N=8$ supergravity
\cite{Cremmer80,Gunaydin:1983rk,Gunaydin:1985cu}.
In table \ref{tab:G/H} the global symmetry group
and its compact subgroup are given for these theories and their
dimensional reductions. In addition we indicate the R-symmetry
group.

\begin{table}[ht]
\begin{center}
\begin{tabular}{||c||c||c|c||c|c||}
\hline \rule[-1mm]{0mm}{6mm}
  $D$ & $H_{\rm R}$ & $G_{\rm pure}$ & $H_{\rm pure}$ & $G_{\rm coupled}$ & $H_{\rm coupled}$ \\
\hline \hline \rule[-1mm]{0mm}{6mm}
  $5$ & $SU(2)$ & $1$ & $1$ & $F_{4}$ & $SU(2) \times Sp(3)$ \\
\hline \rule[-1mm]{0mm}{6mm}
  $4$ & $U(2)$ & $SL(2)$ & $SO(2)$ & $SL(2) \times F_{4}$ & $U(2) \times Sp(3)$ \\
\hline \rule[-1mm]{0mm}{6mm}
  $3$ & $SO(4)$ & $G_2$ & $SO(4)$ & $G_2 \times F_{4}$ & $SU(2) \times SO(4) \times Sp(3)$ \\
\hline
\end{tabular}
\caption{\sl The global symmetries $G$ and their compact subgroups
$H$ of two five-dimensional $\N = 2$ supergravities and their
dimensional reductions. The first is the pure theory while the
second is coupled to seven hyper multiplets. \label{tab:G/H}}
\end{center}
\end{table}

In line with the discussion above, the R-symmetry group is not
contained in $H_{\rm pure}$. This is easy to see in five and four
dimensions. In three dimensions one might think that they coincide.
However, the pure theory reduces to two vector multiplets (and a gravity
multiplet) in three
dimensions. Only one of the $SU(2)$ factors of $H_{\rm R}$ acts on
the scalars contained in the vectors. In contrast, $H_{\rm pure} = SO(4)$ acts
on all 
scalars of the theory. Consequently $H$ and $H_{\rm R}$ should not
be identified; only an $SU(2)$ factor of both groups coincide.
Similarly, in four dimensions only an $SO(2)$ factor coincides.
Therefore the pure theory always has an $SU(2)$ factor of $H_{\rm
R}$ missing in all dimensions, while it has a simple symmetry $G_2$
in three dimensions. In contrast, the theory coupled to hyper
multiplets does have $H_{\rm R} \subseteq  H_{\rm coupled}$, but has
the semi-simple symmetry $G_2 \times F_{4}$ in three dimensions.

We will first discuss the extended semi-simple Lie algebra associated
with the coupled theory before addressing the pure theory.

\subsection{Very-extended $\fg_2 \oplus \ff_4$ and the coupled theory} \label{sec:G2F4}

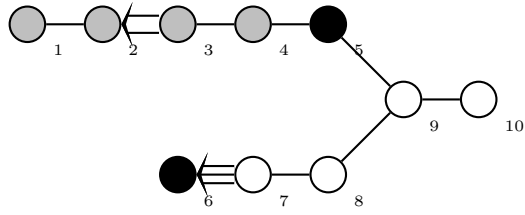
\begin{figure}[h!]
\begin{center}
\begin{pspicture}(2,-1)(8,2)
\dualityNode{2,1}{N1453265943} \nodeLabel{N1453265943}{1}
\dualityNode{3,1}{N2453265943} \nodeLabel{N2453265943}{2}
\dualityNode{4,1}{N3453265943} \nodeLabel{N3453265943}{3}
\dualityNode{5,1}{N4453265943} \nodeLabel{N4453265943}{4}
\disabledNode{6,1}{N5453265943} \nodeLabel{N5453265943}{5}
\disabledNode{4,-1}{N6453265943} \nodeLabel{N6453265943}{6}
\normalNode{5,-1}{N7453265943} \nodeLabel{N7453265943}{7}
\normalNode{6,-1}{N8453265943} \nodeLabel{N8453265943}{8}
\normalNode{7,0}{N9453265943} \nodeLabel{N9453265943}{9}
\normalNode{8,0}{N10453265943} \nodeLabel{N10453265943}{10}
\tripleConnection{N7453265943}{N6453265943}
\singleConnection{N8453265943}{N7453265943}
\doubleConnection{N3453265943}{N2453265943}
\singleConnection{N2453265943}{N1453265943}
\singleConnection{N3453265943}{N4453265943}
\singleConnection{N4453265943}{N5453265943}
\singleConnection{N5453265943}{N9453265943}
\singleConnection{N9453265943}{N10453265943}
\singleConnection{N8453265943}{N9453265943}
\end{pspicture}
\end{center}
\caption{\sl $(\fg_2 \oplus \ff_4)^{+++}$ decomposed as $\fsl(5)^{}
\oplus \ff_{4}^{}$, corresponding to a $D=5$ theory with an $F_4$
internal symmetry group. The colouring in all our diagrams is such that white
nodes correspond to the non-compact gravity line whereas the grey nodes
indicate the internal symmetry group. The black nodes are the ones with
respect to which the level decomposition is performed.} \label{fig:G2F4-5D} 
\end{figure}

The Dynkin diagram of $(\fg_2 \oplus \ff_4)^{+++}$ according to our general
construction of section~\ref{sec:extensions} can be found in 
figure \ref{fig:G2F4-5D}.\footnote{Recall that  $(\fg_2 \oplus \ff_4)^{+++}$
  strictly speaking does not admit a Dynkin diagram, but we refer to the
  diagram of the underlying Kac--Moody algebra as its Dynkin diagram.} A
regular $\fsl(5)$ subalgebra has been 
indicated as well, whose indices will be interpreted as space-time
indices.\footnote{For more details and examples of Kac--Moody decompositions see e.g.~\cite{Kleinschmidt}.} The decomposition into generators of this $\fsl(5)$ with
up to five space-time indices can be
found in table \ref{tab:G2F4-5D} in appendix \ref{sec:tables}. These correspond to the following
physical degrees of freedom (at level $(l_1,l_2)$):
 \begin{itemize}
  \item
   $(0,0)$: the traceless part of the metric, carrying $24$ fields
   of which $10$ will be eliminated due to the local Lorentz symmetry,
  \item
   $(0,0)$: a scalar field which provides the trace of the metric,
  \item
   $(0,0)$: scalars in the adjoint of $\ff_4$, subject to the local
   $SU(2)\times Sp(3)$ symmetry,
  \item
   $(1,0)$: a vector,
  \item
   $(2,0)$: a two-form, which is interpreted as dual to the vector,
  \item
   $(3,0)$: a generator with mixed symmetry properties, which is interpreted
   as the dual graviton,
  \item
   $(0,1)$: three-forms in the adjoint of $\ff_4$, which are interpreted as
   dual to the scalars.
 \end{itemize}
The metric, vector and their duals belong to the graviton multiplet.The
remaining degrees of freedom are spanned by the scalars of the seven hyper
multiplets and their duals. Note that one finds $52$
scalars transforming in the adjoint representation of $\ff_4$, while
there are only $28$ physical scalars in the theory. However, one has
to divide out by the compact subgroup of the internal symmetry group
$F_4$, which is in this case $SU(2) \times Sp(3)$ and therefore
eliminates $24$ scalars. Similarly, one finds $52$ three-forms which
can be seen as the duals to the scalars. We expect that
supersymmetry will impose $24$ linear constraints on the field
strengths of these two-forms. This reduces the number of independent
three-forms to $28$, which will be related to the physical scalars
via a Hodge duality relation. This is completely analogous to the
way the $SL(2) / SO(2)$ sector of IIB supergravity appears in
$E_{11}$ (see e.g.~\cite{Kleinschmidt} and also \cite{IIB-8forms, IIB-10forms}).
In that case the compact $SO(2)$ subgroup
eliminates one of the three scalars. In addition there is a single
linear constraint on the field strengths of the dual eight-forms and
duality relations between the remaining scalars and eight-forms
\cite{IIB-8forms, IIB-10forms}.

Since a crucial aspect of our proposal is that the relevant symmetry is the
quotient of the derived algebra of the Kac--Moody algebra given by the
diagram~\ref{fig:G2F4-5D} we examine the effect of the quotient in this case
in detail. The $(10\times 10)$ Cartan matrix encoded in
figure~\ref{fig:G2F4-5D} has 
one zero eigenvalue so that the associated Kac--Moody algebra must have eleven
independent Cartan generators, including one central charge $\tilde{c}$ and
the associated derivation $\tilde{d}$. The central charge $\tilde{c}$ is the
difference of the central charges of the affine $\mathfrak{f}_4^+$ and
$\mathfrak{g}_2^+$ diagrams contained in fig.~\ref{fig:G2F4-5D}. The
transition to the derived algebra removes $\tilde{d}$ and the quotient
eliminates $\tilde{c}$ so that the resulting algebra has only nine commuting
diagonalisable 
elements. This has to be compared to the number of diagonal metric components
and dilaton-like scalars in the scalar coset. In $D=5$ there are five diagonal
metric components and for the $F_4$ coset there are four dilaton-like scalars
so that the numbers agree precisely.

With this understanding, indeed the above decomposition coincides
exactly with the physical degrees of freedom (and their duals) of
this theory. The other generators do not correspond to any (known)
propagating degrees of freedom. Nevertheless, a subset of these is
known to play an interesting role in supergravity. These are the
purely anti-symmetric $(D-1)$- and $D$-forms, as discussed in the
introduction. In the case at hand these are
 \begin{itemize}
  \item
   $(1,1)$: four-forms in the adjoint of $\ff_4$,
  \item
   $(2,1)$: five-forms in the adjoint of $\ff_4$.
 \end{itemize}
The four-forms are interpreted as dual to the gauge parameters \cite{gaugings1,gaugings2}.
These specify the possible embeddings of a one-dimensional gauge
group in $F_4$. In addition the very-extended
algebra predicts the possibility to include an $\ff_4$ adjoint of
space-time filling five-forms, whose role is yet to be understood.

\begin{figure}[h!]
\begin{center}
\begin{pspicture}(3,-1)(9,2)
\dualityNode{3,1}{N11697693003} \nodeLabel{N11697693003}{1}
\dualityNode{4,1}{N21697693003} \nodeLabel{N21697693003}{2}
\dualityNode{5,1}{N31697693003} \nodeLabel{N31697693003}{3}
\dualityNode{6,1}{N41697693003} \nodeLabel{N41697693003}{4}
\disabledNode{7,1}{N51697693003} \nodeLabel{N51697693003}{5}
\dualityNode{5,-1}{N61697693003} \nodeLabel{N61697693003}{6}
\dualityNode{6,-1}{N71697693003} \nodeLabel{N71697693003}{7}
\disabledNode{7,-1}{N81697693003} \nodeLabel{N81697693003}{8}
\normalNode{8,0}{N91697693003} \nodeLabel{N91697693003}{9}
\normalNode{9,0}{N101697693003} \nodeLabel{N101697693003}{10}
\doubleConnection{N31697693003}{N21697693003}
\singleConnection{N21697693003}{N11697693003}
\singleConnection{N31697693003}{N41697693003}
\singleConnection{N41697693003}{N51697693003}
\tripleConnection{N71697693003}{N61697693003}
\singleConnection{N71697693003}{N81697693003}
\singleConnection{N81697693003}{N91697693003}
\singleConnection{N91697693003}{N101697693003}
\singleConnection{N51697693003}{N91697693003}
\end{pspicture}
\end{center}
\caption{\sl $(\fg_2 \oplus \ff_4)^{+++}$ decomposed as $\fsl(3)^{}
\oplus \fg_{2}^{} \oplus \ff_4$, corresponding to a $D=3$ theory
with a $G_2 \times F_4$ internal symmetry group.}
\label{fig:G2F4-3D}
\end{figure}

In addition to the previous five-dimensional perspective, we would
also like to consider the decomposition of very-extended $\fg_2
\oplus \ff_4$ with respect to its $\fsl(3)$ subalgebra,
corresponding to the theory reduced to three dimensions. The
associated Dynkin diagram is given in figure \ref{fig:G2F4-3D} while
the result of the decomposition can be found in table
\ref{tab:G2F4-3D}. It can easily be seen that at levels
$(l_1,l_2)=(0,0),(0,1)$ and $(1,0)$ one finds exactly the generators
that are associated to the physical degrees of freedom of the
reduced supergravity theory. These are the graviton and its trace,
and scalars in the adjoints of $\fg_2$ and $\ff_4$ and their dual
vectors. In addition the very-extended algebra contains the
following non-propagating degrees of freedom:
 \begin{itemize}
  \item two-forms in the $(\fg_2, \ff_4)$ representations
   $({\bf 1} \oplus {\bf 27},{\bf 1}) \oplus ({\bf 1},{\bf 1} \oplus {\bf
     324}) \oplus ({\bf 14},{\bf 52})$, 
  \item three-forms in the $(\fg_2, \ff_4)$ representations
   $({\bf 1}\oplus {\bf 14}\oplus {\bf 27}\oplus{\bf 77},{\bf 52})
   \oplus({\bf 14},{\bf 1}\oplus{\bf 52}\oplus {\bf 324}\oplus{\bf  1274})$.
 \end{itemize}
The two-forms are in one-to-one correspondence with the components
of the embedding tensor that parametrise the most general gaugings
of this theory \cite{Herger}. Therefore the above features are
exactly what one would expect for the algebra corresponding to this
$D=3$ theory.

Note that in the above we have interpreted the lower branch of the
Dynkin diagram \ref{fig:G2F4-5D} as corresponding to the scalar
manifold of the vectors, while the upper branch corresponds to the
hyper manifolds. In other words, we have chosen to identify
 \begin{align}
  \M_V = \frac{G_2}{SO(4)} \,, \qquad \M_H = \frac{F_4}{SU(2) \times
  Sp(3)} \,,
  \label{3D-identification}
 \end{align}
in three dimensions. With these identifications one can only uplift
to the coupled theories in table \ref{tab:G/H}. In this uplift the
scalar manifold $F_4 / (SU(2) \times  Sp(3))$ is unaffected, while
the scalar manifold $G_2 / SO(4)$ is deconstructed to yield gravity
and vectors in the higher dimensions.

\begin{figure}[h!]
\begin{center}
\begin{pspicture}(3,-1)(9,2)
\dualityNode{5,1}{N1202405117} \nodeLabel{N1202405117}{1}
\dualityNode{6,1}{N2202405117} \nodeLabel{N2202405117}{2}
\disabledNode{7,1}{N3202405117} \nodeLabel{N3202405117}{3}
\dualityNode{3,-1}{N4202405117} \nodeLabel{N4202405117}{4}
\disabledNode{4,-1}{N5202405117} \nodeLabel{N5202405117}{5}
\normalNode{5,-1}{N6202405117} \nodeLabel{N6202405117}{6}
\normalNode{6,-1}{N7202405117} \nodeLabel{N7202405117}{7}
\normalNode{7,-1}{N8202405117} \nodeLabel{N8202405117}{8}
\normalNode{8,0}{N9202405117} \nodeLabel{N9202405117}{9}
\normalNode{9,0}{N10202405117} \nodeLabel{N10202405117}{10}
\singleConnection{N2202405117}{N3202405117}
\singleConnection{N7202405117}{N8202405117}
\singleConnection{N8202405117}{N9202405117}
\singleConnection{N9202405117}{N10202405117}
\singleConnection{N3202405117}{N9202405117}
\tripleConnection{N2202405117}{N1202405117}
\singleConnection{N7202405117}{N6202405117}
\doubleConnection{N6202405117}{N5202405117}
\singleConnection{N5202405117}{N4202405117}
\end{pspicture}
\end{center}
\caption{\sl $(\fg_2 \oplus \ff_4)^{+++}$ decomposed as $\fsl(6)^{}
\oplus \fg_{2}^{} \oplus \fsl(2)$, corresponding to a $D=6$ theory
with a $G_2 \times SL(2)$ internal symmetry group.}
\label{fig:G2F4-6D}
\end{figure}

However, in three dimensions the vector and hyper multiplets are
interchangable, and hence one could have made identification
\eqref{3D-identification} with $\M_V$ and $\M_H$ interchanged. In
this case the scalar manifold $G_2 / SO(4)$ is unaffected by the
uplift, while the other factor $F_4 / (SU(2) \times  Sp(3))$ is
deconstructed in different higher-dimensional fields. With such
identifications the higher-dimensional origin is therefore
completely different. In particular, the highest dimension to which
this theory can be uplifted is six, corresponding to the $\fsl(6)$
regular subalgebra of very-extended $\fg_2 \oplus \ff_4$ shown in
figure \ref{fig:G2F4-6D}. This corresponds to the six-dimensional
chiral $\N = 2$ supergravity that reduces to the $F_4 / (SU(2)
\times Sp(3))$ coset in $D=3$ \cite{Cremmer}, coupled to two hyper
multiplets parametrising a $G_2 / SO(4)$ scalar manifold. We have
checked that the decomposition of very-extended $\fg_2 \oplus \ff_4$
with respect to this $\fsl(6)$ regular subalgebra, given in table
\ref{tab:G2F4-6D}, gives rise to the correct physical degrees of
freedom. In addition it includes the following non-propagating
degrees of freedom:
 \begin{itemize}
  \item five-forms in the $(\fsl(2), \fg_2)$ representations
   $({\bf 2} \oplus {\bf 4}, {\bf 1}) \oplus ({\bf 2},{\bf 14})$,
  \item six-forms in the $(\fsl(2), \fg_2)$ representations
  $({\bf 3} \oplus {\bf 3} \oplus {\bf 5},{\bf 1}) \oplus ({\bf 1} \oplus {\bf
    3},{\bf 14})$.
 \end{itemize}
The former should parametrise the possible gaugings of the $SL(2)
\times G_2$ internal symmetry with the $SL(2)$ doublet of vectors.

\subsection{Very-extended $\fg_2 \oplus \fsu(2)$ and the pure theory} \label{sec:G2A1}

We will now discuss the pure $\N=2$, $D=5$ theory introduced earlier, i.e.~the
case without hyper multiplets, and the corresponding very-extended
algebra.

First let us address in more detail why the case without hyper
multiplets, where the R-symmetry is not contained in $H$, leads to a
mismatch from the Kac--Moody point of view. We will do this in the
context of the $D=5$ pure theory, whose physical bosonic degrees of
freedom reduce in three dimensions to two vector multiplets
parametrising the coset
 \begin{align}
  \M_V = \frac{G_2}{SO(4)} \,.
 \end{align}
The corresponding Kac--Moody algebra is $\fg_2^{+++}$, whose Dynkin
diagram with a regular $\fsl(5)$ subalgebra indicated is given in
figure \ref{fig:G2-5D}. The corresponding decomposition is given in
table \ref{tab:G2-5D} in appendix \ref{sec:tables}, see also
\cite{Kleinschmidt}. At lowest levels
$l$ we find the expected physical degrees of freedom:
 \begin{itemize}
  \item
   $0$: the traceless part of the metric, carrying $24$ degrees of freedom of which $10$ will be eliminated due to the local Lorentz symmetry,
  \item
   $0$: a scalar which provides the trace of the metric,
  \item
   $1$: a vector,
  \item
   $2$: a two-form, which is interpreted as dual to the vector,
  \item
   $3$: a generator with mixed symmetry properties, interpreted as the dual
   graviton.
 \end{itemize}
The other generators all have mixed symmetries and do not correspond
to propagating degrees of freedom.

\begin{figure}[h!]
\begin{center}
\begin{pspicture}(4,0)(8,1)
\disabledNode{4,0}{N11884423562} \nodeLabel{N11884423562}{1}
\normalNode{5,0}{N21884423562} \nodeLabel{N21884423562}{2}
\normalNode{6,0}{N31884423562} \nodeLabel{N31884423562}{3}
\normalNode{7,0}{N41884423562} \nodeLabel{N41884423562}{4}
\normalNode{8,0}{N51884423562} \nodeLabel{N51884423562}{5}
\singleConnection{N41884423562}{N51884423562}
\singleConnection{N41884423562}{N31884423562}
\singleConnection{N31884423562}{N21884423562}
\tripleConnection{N21884423562}{N11884423562}
\end{pspicture}
\end{center}
\caption{\sl $\fg_{2}^{+++}$ decomposed as $\fsl(5)^{}$,
corresponding to a $D=5$ theory.} \label{fig:G2-5D}
\end{figure}
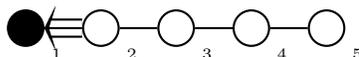

For this particular theory, however, it has been explicitly
calculated which gauge potentials can be included, i.e.~on which
gauge potentials the supersymmetry algebra can be realised
\cite{Gomis}. In addition to the vector and its dual two-form
present in $\fg_2^{+++}$, three- and four-forms transforming under the R-symmetry
$SU(2)$ were found. The four-forms should be expected: they can be
seen as the potentials dual to the gauging parameters. Indeed, in
the pure theory one can gauge a $U(1)$ group \cite{gauged5D}, whose
embedding in $SU(2)$ is described by a triplet of parameters.
However, since the original bosonic fields of the theory (i.e.~the
metric, the vector and their duals) are invariant under the
R-symmetry the corresponding Kac--Moody algebra will not contain
$H_{\rm R}$ and hence will miss the potentials corresponding to its
gauging \cite{Steele}.

A clue for the resolution of this puzzle comes from the triplet of
three-forms. In $D=5$ these would in general be interpreted as dual
to scalars, which the theory does not have and hence would introduce
additional propagating degrees of freedom. However, it was found
that, in order to realise supersymmetry on them, their field
strengths necessarily vanish, and hence they correspond to
non-propagating degrees of freedom. This is similar to the $SL(2)
/ SO(2)$ coset of IIB supergravity (or the $F_4 / (SU(2) \times
Sp(3))$ coset of the coupled theory considered in section
\ref{sec:G2F4}). In the IIB case there was one linear constraint on
the field strengths of the $(D-2)$-forms, corresponding to the local
$SO(2)$ symmetry reducing the number of scalars from three to two.
Therefore a natural interpretation for the three-forms in $D=5$ is
as dual to an $SU(2) / SU(2)$ coset. As this is a compact factor
there are no scalars associated to it, and the dual $(D-2)$-forms
are subject to three linear constraints on their field strengths,
which therefore vanish. Hence the results of \cite{Gomis} indicate
that one should include an $SU(2) / SU(2)$ scalar manifold in the
$D=5$ pure theory (see also \cite{Cremmer80}). This does not mix with the other fields under
dimensional reduction and the resulting coset in $D=3$ is
 \begin{align}
  \M_V = \frac{G_2}{SO(4)} \,, \qquad \M_H = \frac{SU(2)}{SU(2)} \,,
  \label{3D-SU(2)}
 \end{align}
and thus corresponds to a global semi-simple algebra $\fg_2 \oplus
\fsu(2)$.

The same reasoning applies to all $\N = 2$ theories without hyper
multiplets: these have to be extended with an additional compact
$SU(2) / SU(2)$ factor, and therefore reduce to a semi-simple coset
in three dimensions. Very-extended simple algebras like
$\fg_2^{+++}$ should not be associated to an $\N = 2$ theory.
In the conclusions we will discuss whether they correspond
to theories with less than eight supercharges.

\begin{figure}[h!]
\begin{center}
\begin{pspicture}(4,-1)(8,2)
\dualityNode{5,1}{N11042141470} \nodeLabel{N11042141470}{1}
\disabledNode{6,1}{N21042141470} \nodeLabel{N21042141470}{2}
\disabledNode{4,-1}{N31042141470} \nodeLabel{N31042141470}{3}
\normalNode{5,-1}{N41042141470} \nodeLabel{N41042141470}{4}
\normalNode{6,-1}{N51042141470} \nodeLabel{N51042141470}{5}
\normalNode{7,0}{N61042141470} \nodeLabel{N61042141470}{6}
\normalNode{8,0}{N71042141470} \nodeLabel{N71042141470}{7}
\tripleConnection{N41042141470}{N31042141470}
\singleConnection{N41042141470}{N51042141470}
\singleConnection{N21042141470}{N61042141470}
\specialDoubleConnection{N21042141470}{N11042141470}
\singleConnection{N61042141470}{N71042141470}
\singleConnection{N61042141470}{N51042141470}
\end{pspicture}
\end{center}\caption{\sl $(\fg_2 \oplus \fsu(2))^{+++}$ decomposed as $\fsl(5)^{}
\oplus \fsu(2)^{}$, corresponding to a $D=5$ theory with an $SU(2)$
internal symmetry group. Node 1 should be taken compact (see text).}
\label{fig:G2A1-5D}
\end{figure}

Let us now check whether very-extended $(\fg_2 \oplus
\fsu(2))^{+++}$ contains the correct generators. The Dynkin diagram
of this very-extended semi-simple algebra is given in figure
\ref{fig:G2A1-5D} with the $\fsl(5)$ regular subalgebra indicated,
while the corresponding decomposition can be found in table
\ref{tab:G2A1-5D} in appendix \ref{sec:tables}. The additional node
1 should be understood to give rise to an internal $SU(2)$ symmetry
group in $D=5$. Therefore the relevant real form of this algebra is
not maximally non-compact. As the largest part of this paper deals
with algebras of maximally non-compact, or split, real form, we
prefer not introduce the additional notation for other real forms.
This can be found in e.g.~\cite{Keurentjes, Riccioni:2008}. Instead
we will indicate the consequences of the non-split form in what
follows.

A straightforward comparison between the results for $(\fg_2 \oplus
\fsu(2))^{+++}$ and $(\fg_2 \oplus \ff_4)^{+++}$ shows that the
results are completely analogous and will therefore not be listed.
In particular, one finds exactly the same generators which are mentioned
in section \ref{sec:G2F4}, where now instead of the adjoint of
$\ff_4$ one finds the adjoint of $\fsu(2)$. Note that the
corresponding scalars are all pure gauge since the internal symmetry
is compact. In addition there will be three linear constraints on
the field strengths of the dual three-forms. In this way one indeed
recovers the correct physical field content of the graviton
multiplet. Moreover, the three- and four--forms transforming in the
adjoint of $\fsu(2)$ coincide exactly with the results of
\cite{Gomis}. A puzzle arises for the five-forms, which were not
discussed in \cite{Gomis} but are present in the very-extended
algebra. This can be seen as a prediction of our proposal for very-extended semi-simple Lie algebras. As we show in
appendix \ref{sec:five-forms},  it is indeed possible to include an
$SU(2)$-triplet of five-forms as well, provided they have rather
unusual supersymmetry transformations. Hence this example confirms
both the need to include an $SU(2) / SU(2)$ factor and our proposal
for the extended semi-simple algebras.

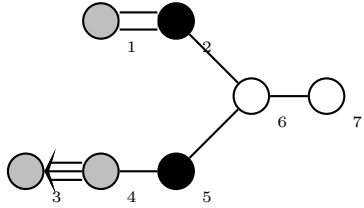
\begin{figure}[h!]
\begin{center}
\begin{pspicture}(4,-1)(8,2)
\dualityNode{5,1}{N12053328494} \nodeLabel{N12053328494}{1}
\disabledNode{6,1}{N22053328494} \nodeLabel{N22053328494}{2}
\dualityNode{4,-1}{N32053328494} \nodeLabel{N32053328494}{3}
\dualityNode{5,-1}{N42053328494} \nodeLabel{N42053328494}{4}
\disabledNode{6,-1}{N52053328494} \nodeLabel{N52053328494}{5}
\normalNode{7,0}{N62053328494} \nodeLabel{N62053328494}{6}
\normalNode{8,0}{N72053328494} \nodeLabel{N72053328494}{7}
\singleConnection{N62053328494}{N72053328494}
\singleConnection{N62053328494}{N52053328494}
\singleConnection{N52053328494}{N42053328494}
\tripleConnection{N42053328494}{N32053328494}
\specialDoubleConnection{N12053328494}{N22053328494}
\singleConnection{N22053328494}{N62053328494}
\end{pspicture}
\end{center}
\caption{\sl $(\fg_2 \oplus \fsu(2))^{+++}$ decomposed as
$\fsl(3)^{} \oplus \fg_2 \oplus \fsu(2)^{}$, corresponding to a
$D=3$ theory with an $G_2 \times SU(2)$ internal symmetry group.
Node 1 should be taken compact (see text).} \label{fig:G2A1-3D}
\end{figure}

Again we also consider the three-dimensional theory associated to
the very-extended algebra. The corresponding Dynkin diagram is given
in figure \ref{fig:G2A1-3D} while the decomposition can be found in
table \ref{tab:G2A1-3D}. It can be verified that this gives rise to
the correct physical degrees of freedom: the graviton and scalars in
the adjoints of $\fg_2$ and $\fsu(2)$, and their duals. In addition
the very-extended algebra contains the following two- and
three-forms:
 \begin{itemize}
  \item two-forms in the $(\fg_2, \fsu(2))$ representations
   $({\bf 1} \oplus {\bf 1} \oplus{\bf 27},{\bf 1}) \oplus ({\bf 14},{\bf 3})$,
  \item three-forms in the $(\fg_2, \fsu(2))$ representations
  $({\bf 14} \oplus{\bf 14} \oplus{\bf 27} \oplus{\bf 64},{\bf 1}) \oplus
  ({\bf 1} \oplus{\bf 14} \oplus {\bf 14} \oplus {\bf 27} \oplus{\bf 77},{\bf
    3} )$.
 \end{itemize}
The two-forms in this algebra are in perfect agreement with the
components of the embedding tensor and thus with the possible
gaugings \cite{Herger}. This is a further confirmation of our
proposal.

Note that, in contrast to the $(\fg_2 \oplus \ff_4)^{+++}$ case
discussed in the previous section, the present theory does not allow
for an alternative uplift. In other words, if we would interchange
the vector and hyper multiplets in \eqref{3D-SU(2)} then the theory
does not allow for an uplift to $D \geq 4$. The reason is that the
scalars of $\M_V$ have to provide the degrees of freedom for
e.g.~the metric in higher dimensions. The scalar manifold $SU(2) /
SU(2)$, however, does not have any degrees of freedom associated to
it. Therefore with this interpretation the theory only lives in
$D=3$. From the point of view of the Dynkin diagram in figure
\ref{fig:G2A1-5D} one might think that there is an alternative
$\fsl(4)$ regular subalgebra that includes the affine extension node
of $\fsu(2)$, but this is not possible due to the non-split form; in the theory of group oxidation,
the $\fsl(D)$ regular subalgebra is not allowed to be
connected to the compact node 1 \cite{Keurentjes}.

\section{Relation to $E_{11}$} \label{sec:E11}

In this section we show that our proposal is consistent with the
possibility to obtain certain $\N \le 2$ theories as truncation of
the maximal theory. In particular we will focus on the pure and
coupled $\N = 2$ theories, both of which have an $\N=8$ origin
\cite{Cremmer80,Gunaydin:1983rk,Gunaydin:1985cu}. We assume that maximal theories are
described by an $E_8^{+++}\equiv E_{11}$ symmetry (in split form)
and will verify that the quotients discussed in the preceding
sections arise as subalgebras of $\mathfrak{e}_{11}$. In order to
make the analysis rigorous we first review some facts about
$\fe_{11}$ and then move on to proving $(\fg_2 \oplus
\ff_4)^{+++}\subseteq \mathfrak{e}_{11}$ and $(\fg_2 \oplus
\fsu(2))^{+++}\subseteq \mathfrak{e}_{11}$.

\subsection{$\mathfrak{e}_{11}$ in $\mathfrak{sl}(11)$ basis}
\label{sec:e11a10}

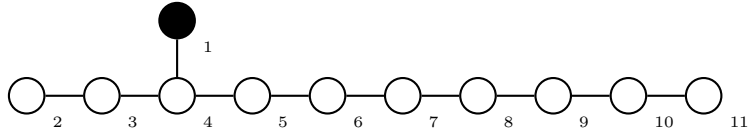
\begin{figure}[h!]
\begin{center}
\begin{pspicture}(2,0)(11,2)
\disabledNode{4,1}{N11723083412}
\nodeLabel{N11723083412}{1}
\normalNode{2,0}{N21723083412}
\nodeLabel{N21723083412}{2}
\normalNode{3,0}{N31723083412}
\nodeLabel{N31723083412}{3}
\normalNode{4,0}{N41723083412}
\nodeLabel{N41723083412}{4}
\normalNode{5,0}{N51723083412}
\nodeLabel{N51723083412}{5}
\normalNode{6,0}{N61723083412}
\nodeLabel{N61723083412}{6}
\normalNode{7,0}{N71723083412}
\nodeLabel{N71723083412}{7}
\normalNode{8,0}{N81723083412}
\nodeLabel{N81723083412}{8}
\normalNode{9,0}{N91723083412}
\nodeLabel{N91723083412}{9}
\normalNode{10,0}{N101723083412}
\nodeLabel{N101723083412}{10}
\normalNode{11,0}{N111723083412}
\nodeLabel{N111723083412}{11}
\singleConnection{N11723083412}{N41723083412}
\singleConnection{N41723083412}{N31723083412}
\singleConnection{N31723083412}{N21723083412}
\singleConnection{N41723083412}{N51723083412}
\singleConnection{N51723083412}{N61723083412}
\singleConnection{N61723083412}{N71723083412}
\singleConnection{N71723083412}{N81723083412}
\singleConnection{N81723083412}{N91723083412}
\singleConnection{N91723083412}{N101723083412}
\singleConnection{N101723083412}{N111723083412}
\end{pspicture}
\end{center}
\caption{\sl $\fe_{11}$ decomposed as
$\fsl(11)^{}$, corresponding to the $D=11$ maximal theory.} \label{fig:E11-11D}
\end{figure}

The Dynkin diagram of $\mathfrak{e}_{11}$ is shown in fig.~\ref{fig:E11-11D}
where the exceptional node has been marked as deleted so that there is an
$\mathfrak{sl}(11)$ gravity line corresponding to a theory in $D=11$. At level
$(l)$ the spectrum contains \cite{West1, damourhenneauxnicolai}:
 \begin{itemize}
  \item
   $(0)$: the traceless part of the metric, carrying $120$ fields
   of which $55$ will be eliminated due to the local Lorentz symmetry,
  \item
   $(0)$: a scalar which provides the trace of the metric and turning
   $\mathfrak{sl}(11)$ into $\mathfrak{gl}(11)$,
  \item
   $(1)$: a three-form as present in $D=11$ supergravity,
  \item
   $(2)$: a six-form, which is interpreted as dual to the three-form,
  \item
   $(3)$: a generator with mixes properties, interpreted as the dual graviton.
 \end{itemize}

We will use the following notation for the generators on levels $(0)$ and $(1)$:
\begin{eqnarray}
\mathfrak{gl}(11) &:& K^a{}_b \,,\quad a,b=1,\ldots,11\,,\nonumber\\
\text{three-form} &:& E^{a_1a_2a_3}=E^{[a_1a_2a_3]}\,.
\end{eqnarray}
They commute according to \cite{West1,damourhenneauxnicolai}
\begin{eqnarray}
\left[K^a{}_b, K^c{}_d\right] &=& \delta^c_b K^a{}_d - \delta^a_d
  K^c{}_d\,,\nonumber\\
\left[K^a{}_b, E^{c_1c_2c_3}\right] &=&
  3\delta_b^{[c_1}E^{c_2c_3]a}\,.
\end{eqnarray}

The Lie algebra $\mathfrak{e}_{11}$ is defined in terms of simple Chevalley
generators $e_i$, $f_i$ and $h_i$ for $i=1,\ldots,11$ with the relations
\cite{Kac}
\begin{align}\label{eq:csrels}
\left[h_i,h_j\right] &= 0\,,&\quad
   \left[e_i,f_j\right] &= \delta_{ij} h_j\,,&\nonumber\\
\left[h_i,e_j\right] &= A_{ij} e_j \,,&\quad
   \left[h_i,f_j\right] &= -A_{ij}f_j\,,&\nonumber\\
(\text{ad}\,e_i)^{1-A_{ij}}e_j &=0 \,,&\quad
   (\text{ad}\,f_i)^{1-A_{ij}}f_j &=0 \,.&
\end{align}
where $A_{ij}$ are the entries of the Cartan matrix of
$\mathfrak{e}_{11}$. The generators $h_i$ and $e_i$ can be chosen to be
related to the basis we chose above by \cite{West1,damournicolai}
\begin{eqnarray}\label{eq:E11gen}
h_i &=& K^{13-i}{}_{13-i}- K^{12-i}{}_{12-i}
   \quad\quad\text{(for $i=2,\ldots,11$)}\,,\nonumber\\
h_{1} &=& -\frac13\sum_{a=1}^{11} K^a{}_a
   + K^9{}_9 + K^{10}{}_{10} + K^{11}{}_{11}\,,\nonumber\\
e_i &=& K^{13-i}{}_{12-i}
   \quad\quad\text{(for $i=2,\ldots,11$)}\,,\quad
e_{1} = E^{9\,10\,11} \,,\nonumber\\[1mm]
f_i &=& K^{12-i}{}_{13-i}
   \quad\quad\text{(for $i=2,\ldots,11$)}\,,\quad
f_{1} = F_{9\,10\,11} = (E^{9\,10\,11})^T \,.
\end{eqnarray}

\subsection{$(\mathfrak{g}_2\oplus
  \mathfrak{f}_4)^{+++}\subset\mathfrak{e}_{11}$} \label{sec:g2f4e11}

To show that $(\mathfrak{g}_2\oplus \mathfrak{f}_4)^{+++}$ as defined in
section~\ref{sec:extensions} is a subalgebra of $\mathfrak{e}_{11}$ we exhibit
defining Chevalley generators $H_i$ and $E_i$ (for $i=1,\ldots,10$) as
combinations of $\mathfrak{e}_{11}$ generators (\ref{eq:E11gen}) such that
they obey the relations (\ref{eq:csrels}) but now with the Cartan matrix
$A_{ij}$ encoded in
diagram~\ref{fig:G2F4-5D}. This will describe the derived algebra needed in
the construction of $(\mathfrak{g}_2\oplus \mathfrak{f}_4)^{+++}$; the
quotient by the centre will follow from the fact that the central element
corresponds to the zero element of $\mathfrak{e}_{11}$ and therefore is not a
linearly independent Cartan generator.

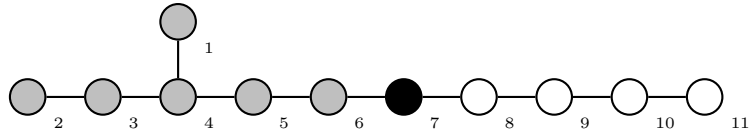
\begin{figure}[h!]
\begin{center}
\begin{pspicture}(2,0)(11,2)
\dualityNode{4,1}{N11083197328}
\nodeLabel{N11083197328}{1}
\dualityNode{2,0}{N21083197328}
\nodeLabel{N21083197328}{2}
\dualityNode{3,0}{N31083197328}
\nodeLabel{N31083197328}{3}
\dualityNode{4,0}{N41083197328}
\nodeLabel{N41083197328}{4}
\dualityNode{5,0}{N51083197328}
\nodeLabel{N51083197328}{5}
\dualityNode{6,0}{N61083197328}
\nodeLabel{N61083197328}{6}
\disabledNode{7,0}{N71083197328}
\nodeLabel{N71083197328}{7}
\normalNode{8,0}{N81083197328}
\nodeLabel{N81083197328}{8}
\normalNode{9,0}{N91083197328}
\nodeLabel{N91083197328}{9}
\normalNode{10,0}{N101083197328}
\nodeLabel{N101083197328}{10}
\normalNode{11,0}{N111083197328}
\nodeLabel{N111083197328}{11}
\singleConnection{N11083197328}{N41083197328}
\singleConnection{N41083197328}{N31083197328}
\singleConnection{N31083197328}{N21083197328}
\singleConnection{N41083197328}{N51083197328}
\singleConnection{N51083197328}{N61083197328}
\singleConnection{N61083197328}{N71083197328}
\singleConnection{N71083197328}{N81083197328}
\singleConnection{N81083197328}{N91083197328}
\singleConnection{N91083197328}{N101083197328}
\singleConnection{N101083197328}{N111083197328}
\end{pspicture}
\end{center}
\caption{\sl $\fe_{11}$ decomposed as $\fsl(5)^{} \oplus \fe_6$, corresponding to the $D=5$ maximal
theory with an $E_6$ internal symmetry group.} \label{fig:E11-5D}
\end{figure}

In order to obtain diagram~\ref{fig:G2F4-5D} it is most convenient
to also take the $D=5$ version of the $\mathfrak{e}_{11}$ diagram.
This is shown in figure~\ref{fig:E11-5D}, where the global $E_{6}$
symmetry of the scalar manifold of maximal ungauged supergravity in
$D=5$ is evident. The $F_{4}$ symmetry describing the scalar of the
truncated coupled theory discussed in section~\ref{sec:G2F4} is the maximal
subgroup $F_{4}\subset E_{6}$ whose defining Lie algebra generators
we will give explicitly below. In addition we need to find an $F_4$
singlet vector generator which gives node $6$ of
diagram~\ref{fig:G2F4-5D} and an $F_4$ adjoint three-form which
gives node $5$ of that diagram.

The correct choices for the Cartan generators are
\begin{eqnarray}\label{eq:newcsi}
H_{10} = h_{11} \,,\quad H_{9} = h_{10}\,,\quad
H_8 = h_9 \,,\quad H_7 =h_8\,,
\end{eqnarray}
since the gravity generators are common in both theories,
\begin{eqnarray}
H_4 = h_{1}\,,\quad H_3 = h_4\,,\quad
 H_2 = h_3 + h_5 \,,\quad H_1 = h_2 + h_{6}\,,
\end{eqnarray}
since this is the correct embedding of $\mathfrak{f}_4\subset\mathfrak{e}_6$
and
\begin{eqnarray}\label{eq:newcsm}
H_6 &=& 3h_7 + 4h_6 + 5h_5 + 6h_4 + 4h_3 + 2h_{2} + 3 h_{1}\nonumber\\
    &=& - K^1{}_1 - K^2{}_2 - K^3{}_3 - K^4{}_4 + 2K^5{}_5\,,\nonumber\\
H_5 &=& h_9 + 2 h_8 + 3h_7 + 3h_6 + 3h_5 + 3h_4 + 2h_3 + h_{2} +
            h_{1}\nonumber\\
    &=& -\frac13\sum_{a=1}^{11} K^a{}_a + K^3{}_3 + K^4{}_4 + K^5{}_5
\end{eqnarray}
for the connecting nodes. This choice becomes clearer with the expression for
the simple positive step operators
\begin{eqnarray}
E_{10} = e_{11} \,,\quad E_{9} = e_{10}\,,\quad
E_8 = e_9 \,,\quad E_7 = e_8\,,
\end{eqnarray}
again since the gravity sectors are common in both theories,
\begin{eqnarray}
E_4 = e_{1}\,,\quad E_3 = e_4\,,\quad
 E_2 = e_3 + e_5 = K^7{}_8 + K^9{}_{10}  \,,\quad
 E_1 = e_2 + e_{6} = K^6{}_7 + K^{10}{}_{11}\,,
\end{eqnarray}
from the embedding of $\mathfrak{f}_4\subset\mathfrak{e}_6$ and
\begin{eqnarray}\label{eq:newcsf}
E_6 &=& E^{5\,6\,11} - E^{5\,7\,10} + E^{5\,8\,9}\,,\nonumber\\
E_5 &=& E^{3\,4\,5}\,.
\end{eqnarray}
The generators $F_i$ are the transposes of the $E_i$ given here.

With the formul\ae{} (\ref{eq:newcsi})--(\ref{eq:newcsf}) one verifies
the relations (\ref{eq:csrels}), for example
\begin{eqnarray}
\left[H_6,E_6\right] = 2 E_6\,,\quad
\left[H_7,E_6\right] = - E_6\,,\quad
\left[H_6, E_7\right] = - 3E_7\,,\,\text{etc.}
\end{eqnarray}
For the verification of the Serre relations it is useful to notice that all
simple step operators belong $\mathfrak{sl}(2)$ subalgebras and the
computations can be shortened by using the representation theory of
$\mathfrak{sl}(2)$; otherwise some of the Serre relations involve
$\mathfrak{e}_{11}$ commutators up to $\mathfrak{sl}(11)$ level $l=5$ which
exceeds the level to which the relations have been worked out.\footnote{The
  highest known commutators for $\mathfrak{e}_{10}$ involve level $l=4$
  \cite{Fischbacher:2005fy}.}

An important observation regarding the new Cartan generators $H_i$ of
(\ref{eq:newcsi})--(\ref{eq:newcsm}) is that they are not all linearly
independent: The combinations
\begin{eqnarray}\label{eq:G2F4centres}
c_{\fg_2} &=& H_6 + 2 H_7 + H_8 \nonumber\\
       &=& h_9 + 2 h_8 + 3h_7 + 4h_6 + 5h_5 + 6h_4 + 4h_3 + 2h_{2} + 3
       h_{3}\,,\nonumber\\
c_{\ff_4} &=& H_1 + 2H_2 + 3H_3 + 2H_4 + H_5 \nonumber\\
       &=&  h_9 + 2 h_8 + 3h_7 + 4h_6 + 5h_5 + 6h_4 + 4h_3 + 2h_{2} + 3
       h_{3}\,
\end{eqnarray}
are identical so that $c_{\fg_2}-c_{\ff_4}=0$.\footnote{We note also that both are
  identical to the central element of
  $\mathfrak{e}_8^+\subset\mathfrak{e}_{11}$.}
The combinations above were chosen
because $c_{\fg_2}$ is the central charge of $\mathfrak{g}_2^+$ and $c_{\ff_4}$
that of $\mathfrak{f}_4^+$ contained in the diagram. Therefore,
$c_{\fg_2}-c_{\ff_4}$ is precisely the combination that has to be quotiented out
of the abstract (derived) algebra generated by $H_i$, $E_i$ and $F_i$
according to our definition of section~\ref{sec:extensions}. This is realised
here automatically in $\mathfrak{e}_{11}$ since this combination corresponds
to the zero element in $\mathfrak{e}_{11}$. Hence we have shown that the
quotient Lie algebra $(\mathfrak{g}_2\oplus \mathfrak{f}_4)^{+++}$ is a
subalgebra of the Kac--Moody algebra $\mathfrak{e}_{11}$.Therefore
any $E_{11}$ invariant dynamics describing $D=11$
supergravity (or an extension thereof) would entail a consistent truncation
to the $\N=2$ theory with seven
hypermultiplets invariant under the algebra described in
section~\ref{sec:G2F4}.

The definitions (\ref{eq:newcsi})--(\ref{eq:newcsf}) also show the $D=11$ origin of
the fields of the $\N=2$ model in $D=5$: The vector of the gravity multiplet
corresponds to the $F_4$ invariant combination of the $27$ vector fields one
obtains from the three-form of $D=11$, whereas the scalars are obtained by the
truncation of the $E_6$ scalar coset to the $F_4$ scalar coset.

\subsection{$(\mathfrak{g}_2\oplus
  \mathfrak{su}(2))^{+++}\subset\mathfrak{e}_{11}$} \label{sec:g2a1e11}

Here we show that the quotient algebra $(\mathfrak{g}_2\oplus
  \mathfrak{su}(2))^{+++}$, which was discussed in section~\ref{sec:G2A1} in
  relation to pure and simple $\N=2$ supergravity in $D=5$,  is a subalgebra
  of $\mathfrak{e}_{11}$. This can be seen most easily by using the result of
  the preceding section that $(\ff_4\oplus \fg_2)^{+++}$ is a subalgebra of
  $\mathfrak{e}_{11}$ and embedding $(\mathfrak{g}_2\oplus 
  \mathfrak{su}(2))^{+++}$ in $(\ff_4\oplus \fg_2)^{+++}$.
  
We recall that the local part of the internal symmetries of the coupled theory
is $Sp(3)\times SU(2)$, the compact subgroup of $F_4$ (in split form). The
$SU(2)$ factor is the R-symmetry group in $D=5$. It is natural to decompose
all $\ff_4$ representations, which appear in the level decomposition of
$(\ff_4\oplus \fg_2)^{+++}$, under
$\mathfrak{sp}(3)\oplus\mathfrak{su}(2)$. Inspection of
table~\ref{tab:G2F4-5D} shows that the only $\ff_4$ representations occurring
for fields with at most five space-time indices are the ${\bf 1}$ and the
${\bf 52}$ of $\ff_4$. They decompose as 
\begin{eqnarray}\label{eq:f4toc1c3} 
{\bf 1} &\rightarrow& ({\bf 1},{\bf 1})\nonumber\\
{\bf 52} &\rightarrow& ({\bf 1},{\bf 3}) \oplus ({\bf 14}_s, {\bf 2}) \oplus ({\bf 21},{\bf 1})
\end{eqnarray}
under $\mathfrak{sp}(3)\oplus\mathfrak{su}(2)\subset \ff_4$. The restriction
to $\mathfrak{sp}(3)$ singlets leaves only $\mathfrak{su}(2)$ singlets and
triplets. Performing this restriction to $\mathfrak{sp}(3)$ invariant
generators within $(\ff_4\oplus \fg_2)^{+++}$ on the first levels yields
precisely the fields with at most five space-time indices transforming under
$\mathfrak{su}(2)$ as those of table~\ref{tab:G2A1-5D}, which lists the lowest
levels of $(\mathfrak{g}_2\oplus  \mathfrak{su}(2))^{+++}$ decomposed with
respect to $\mathfrak{sl}(5)\oplus\mathfrak{su}(2)$.  

The restriction to $\mathfrak{sp}(3)$ invariant states in
$(\ff_4\oplus\fg_2)^{+++}$ defines a subalgebra $\mathfrak{s}$. Since the
fundamental generators of $(\mathfrak{g}_2\oplus  \mathfrak{su}(2))^{+++}$ on
levels $(0,1)$ and $(1,0)$ are contained in this subalgebra, we deduce that
$(\mathfrak{g}_2\oplus  \mathfrak{su}(2))^{+++}\subset
(\ff_4\oplus\fg_2)^{+++}\subset\mathfrak{e}_{11}$.\footnote{That the inclusion
  relation does not only hold at the level of generators but also at the level
  of Lie brackets follows from the Serre relations which can be deduced for
  the invariant generators from the embedding.} The embedding into
$\mathfrak{e}_{11}$ could also be carried out directly by decomposing all
generators of $\mathfrak{e}_{11}$ under
$\mathfrak{sl}(5)\oplus\mathfrak{sp}(3)\oplus\mathfrak{su}(2)$ and then
restricting to $\mathfrak{sp}(3)$ singlets. For the fields with up to five
space-time indices the invariant generators are identical to those of
$(\mathfrak{g}_2\oplus  \mathfrak{su}(2))^{+++}$ but more branchings of the
type (\ref{eq:f4toc1c3}) are required to make this manifest, giving a less
transparent derivation. 

It is of  interest to ask whether the subalgebra $\mathfrak{s}$ of
$(\ff_4\oplus\fg_2)^{+++}$ defined by $\mathfrak{sp}(3)$ invariance is
identical to $(\mathfrak{g}_2\oplus  \mathfrak{su}(2))^{+++}$. For the fields
with up to five space-time indices there is no difference but one can check by
computer analysis that $(\mathfrak{g}_2\oplus  \mathfrak{su}(2))^{+++}$ is in
fact a proper subalgebra of $\mathfrak{s}$. By virtue of general
characterisation theorems the complexified subalgebra $\mathfrak{s}$ is a
Borcherds algebra \cite{Borcherds:1988}.\footnote{This is similar to the way
  in which the pure $\N=4$ algebra $\mathfrak{d}_8^{+++}$ is contained in a
  Borcherds subalgebra of $\mathfrak{e}_{11}$ \cite{Kleinschmidt:2004dy}.
  There the Borcherds algebra can be constructed by keeping only tensor
  representations of a $\mathfrak{d}_{10}$ common to $\mathfrak{e}_{11}$ and
  $\mathfrak{d}_8^{+++}$ whereas here we have a stronger restriction to
  $\mathfrak{sp}(3)$ invariant tensors.} Since to date generally only
generators with at most $D$ indices in a decomposition under a
$\mathfrak{sl}(D)$ subalgebra have a supergravity interpretation, we
cannot offer an explanation of this difference here (see
\cite{Riccioni:2006az} for a discussion of this point).

The real form of $(\mathfrak{g}_2\oplus  \mathfrak{su}(2))^{+++}$ follows also
from the embedding above:  Because the scalars at level $(0,0)$ inside the
split $(\ff_4\oplus\fg_2)^{+++}$ which are invariant under $\mathfrak{sp}(3)$
all belong to the compact $\mathfrak{su}(2)\subset \ff_4$, it follows that the
$A_1$ node of the diagram of $(\mathfrak{g}_2\oplus  \mathfrak{su}(2))^{+++}$
has to be compact so that the associated summand in the extension process is
the compact $\mathfrak{su}(2)$. 

The construction of $(\mathfrak{g}_2\oplus  \mathfrak{su}(2))^{+++}$ inside
$(\ff_4\oplus\fg_2)^{+++}\subset \mathfrak{e}_{11}$ also gives a $D=11$ origin
to the propagating fields of pure $\N=2$ supergravity in $D=5$ as well as of
its possible deformations.

We expect that a similar analysis of the quotient algebra can be carried out
in all cases when the $\N\le 2$ theory has a $D=11$ origin.

\section{Discussion} \label{sec:discussion}

In this paper we have discussed the extended semi-simple symmetries
that play a role in $\N \leq 2$ supergravity. In particular, we have
put forward a proposal for the extension of semi-simple Lie
algebras, corresponding to the well-known Kac--Moody extensions of
simple Lie algebras and based on analogous reasoning leading to the
affine extension as the Geroch group. In addition we have argued
that an $SU(2) / SU(2)$ scalar manifold has to be coupled to all $\N
= 2$ theories without hyper multiplets, such that these fall in the
semi-simple realm as well. Support for these conjectures has been
gathered from a number of different points of view:
\begin{itemize}
 \item
 The extended semi-simple Lie algebras give rise to the correct physical degrees of freedom in the $\N=2$ supergravity examples discussed.
 \item
 They contain the correct generators corresponding to the non-propagating deformation potentials. In particular, very-extended $\fg_2 \oplus \ff_4$ contains the correct deformation potentials for the coupled theory in $D=5$, corresponding to the gauging of a single isometry of the scalar
 manifold of the hyper multiplets. Very-extended $\fg_2 \oplus
 \fsu(2)$ reproduces the results of \cite{Gomis} on the
 deformation potentials of the pure theory in $D=5$, corresponding to the gauging of a $U(1) \subset SU(2)$ of the R-symmetry group. Finally, both very-extended algebras
 contain the correct deformation potentials in $D=3$, where the gaugings have been
 classified in \cite{Herger}.
 \item
 Very-extended $\fg_2 \oplus \fsu(2)$ predicts the possibility to include
 a triplet of top-forms in $D=5$, which is checked successfully in appendix \ref{sec:five-forms}.
 The supersymmetry variations of these five-forms seem to be of a
 novel type.
 \item
 Both very-extended (quotient) algebras are subalgebras of $\fe_{11}$,
 corresponding to the truncation of $\N=8$ supergravity to the associated $\N =2$ theories.
\end{itemize}
We would like to emphasise that our results can be applied to all semi-simple $D=3$ cosets. As we have argued, the homogeneous scalar manifolds will generically be semi-simple with two simple factors for $\N=2$ supergravity. In addition, the simple algebras have to be augmented with an $\fsu(2)$ factor.

An additional illustration of the latter is provided by pure $\N = 2$ supergravity in four dimensions. Upon reduction to three dimensions this gives rise to the coset
 \begin{align}
  \M_V = \frac{SU(2,1)}{S(U(2) \times U(1))} \,.
 \end{align}
The point is that $\fsu(2,1)^{+++}$ does not contain any three-forms
when decomposed with respect to $SL(4)$ and hence predicts no
gaugings of this four-dimensional supergravity. However, it is known
that one can gauge a $U(1)$ in the $SU(2)$ part of the R-symmetry
group \cite{Freedman, Fradkin}. One can not gauge the separate
$U(1)$ factor of $H_{\rm R} = SU(2) \times U(1)$ as the vector and
its Hodge dual transform as a doublet under it. We have checked that
very-extended $\fsu(2,1) \oplus \fsu(2)$ contains deformation
potentials transforming as an $SU(2)$ triplet and a $U(1)$ doublet,
consistent with this gauging. In addition it predicts an $SU(2)$
triplet of two-forms with vanishing field strengths and two $SU(2)$
triplets of four-forms in this theory. It would be interesting to
include these explicitly, similar to \cite{Gomis}.

We expect the same phenomenon of extended semi-simple Lie algebras to play a role in $\N=1$ supergravities. These theories reduce to a K\"{a}hler scalar
manifold\footnote{The additional topological constraint requiring it
to be Hodge-K\"{a}hler \cite{Bagger-Witten} is not important here.}
in three dimensions \cite{Zumino, Bagger}, homogeneous examples of which
can be found in e.g.~\cite{VanProeyen:2003}. An
important point is that the product of two K\"{a}hler manifolds is
again a K\"{a}hler manifold. Therefore one can couple any number of
homogeneous spaces to $\N = 1$ supergravity. Such products appear
naturally in e.g.~the truncation from $\N=2$ to $\N=1$ \cite{Ferrara}. Hence, for $\N=1$ supergravity, in three dimensions one can have semi-simple algebras of isometries with more than two
factors: $\fg_a \oplus \fg_b \oplus \fg_c \oplus \ldots$. The
extensions of such semi-simple algebras were outlined in section
\ref{sec:extensions}.

Due to the sum of simple factors in $\fg$, one can argue that the
algebra of isometries of homogeneous scalar manifolds is generically
semi-simple for $\N=1$ supergravity. This is similar to what we
found for $\N=2$ supergravity. However, the $\N=2$ situation is
different in that one can not add any number of homogeneous spaces,
as the product of quaternionic-K\"{a}hler spaces is not
quaternionic-K\"{a}hler itself (except when they are
hyper-K\"{a}hler spaces, which are not allowed in $\N=2$
supergravity). Hence one can only couple two such spaces, associated
with $\N =2$ vector and hyper multiplets.

An example of a semi-simple algebra appears in $\N=1$, $D=4$ supergravity coupled
to a chiral multiplet. After reduction to three dimensions this theory corresponds to the coset
 \begin{align}
  \frac GH = \frac{SL(2)}{SO(2)} \times \frac{SL(2)}{SO(2)} \,.
 \end{align}
In line with the previous discussion it consists of a product of K\"{a}hler manifolds. The hidden symmetry group is therefore semi-simple: $SL(2) \times SL(2) \simeq SO(2,2)$. This theory and the very-extended algebra $(\fsl(2) \oplus \fsl(2))^{+++}$ are discussed in detail in appendix \ref{sec:N=1}. Based on covariance
with respect to $\fso(2,2+n_{\rm V})$, we confirm that the very-extended semi-simple algebra
$(\fsl(2) \oplus \fsl(2))^{+++}$ contains exactly the generators one would expect.

However, in addition to the appearance of semi-simple symmetries, we also expect a number of new features to appear for $\N = 1$. This can be seen from the three-dimensional classification of \cite{Herger}. A first point is that one has to incorporate the $SO(2)$ R-symmetry group as a central extension of the isometry group. Furthermore, in contrast to $\N \geq 2$, there is no linear constraint on the possible gaugings for $\N = 1$. Finally, there are deformations that do not correspond to a gauging but rather to the addition of a superpotential. Therefore we would not expect e.g.~$(\fsl(2) \oplus \fsl(2))^{+++}$ to describe all the deformations of the $\N=1$ theory. It remains to be seen to what extent all $\N=1$ features can be reproduced from the Kac--Moody side.

Similarly, one could consider supergravity theories with six supersymmetries. These theories have a single quaternionic-K\"{a}hler manifold in three dimensions \cite{Tollsten}.
The R-symmetry group is $H_{\rm R} = SO(3)$ and hence these do not suffer from the
R-symmetry problems discussed. Therefore one could expect that
very-extended simple Lie algebras, such as $\fg_2^{+++}$, incorporate both the physical and the
non-propagating degrees of freedom of this theory. However, it turns out that for these theories
there is no linear constraint on the possible gaugings either \cite{Herger} and hence the Kac--Moody correspondence remains unclear as well.

Finally, the discussion so far has been concerned with the correspondence between
supergravity and (quotients of) Kac--Moody algebras to $\N \leq 2$
theories with homogeneous spaces. More specifically, we have
restricted ourselves to homogeneous scalar manifolds with
semi-simple groups of isometries. Interesting future research would
be to investigate the extended algebras associated to homogeneous
but non-semi-simple groups, or even to venture into the realm of
non-homogeneous scalar manifolds.

Besides these interesting points regarding the correspondence for the bosonic
sectors of various supergravity theories it would be worthwhile to extend our
new cases also to the fermionic sector. It is known that the maximally
supersymmetric theories in $D=11$ and $D=10$ have propagating fermionic
degress of freedom which can be grouped into finite-dimensional (unfaithful)
representations of $K(E_{11})$ \cite{Kleinschmidt:2006tm,Damour:2005zs,de
  Buyl:2005mt} and also the half-maximal case has been analysed
\cite{Hillmann:2006ic}. We strongly expect that the compact subalgebras of the
quotient algebras presented here will also admit finite-dimensional spinor
representations which correspond to the fermionic degrees of freedom of the
various $\N\leq 2$ theories they belong to.

\section*{Acknowledgements}

We thank Eric Bergshoeff, Alessio Celi, Alex Feingold, Joaquim Gomis, Marc
Henneaux, Hermann Nicolai, Teake Nutma, Henning Samtleben, Igor
Schnakenburg and Antoine van Proeyen for very stimulating
discussions. In addition, we acknowledge using the SimpLie program
of Teake Nutma for the decompositions of appendix \ref{sec:tables}.
The authors would like to thank each other's institutions for
hospitality. A.K. is a Research Associate of the Fonds de la
Recherche Scientifique--FNRS, Belgium. The work of D.R.~has been
supported by the European EC-RTN project MRTN-CT-2004-005104, MCYT
FPA 2004-04582-C02-01 and CIRIT GC 2005SGR-00564.

\appendix

\section{Five-forms in pure and simple $D=5$ supergravity} \label{sec:five-forms}

In \cite{Gomis} it has been analysed on which higher-rank potentials
the supersymmetry algebra of pure and simple $D=5$ supergravity can
be analysed. We will briefly review their results for the ungauged
case and then show that it is possible to introduce a further
extension. All conventions are identical to \cite{Gomis} where more
details can be found.

In the standard formulation the supersymmetry algebra is realised on
the metric, a gravitino and a vector. Their supersymmetry
transformations are given by
 \begin{align}
  \delta e_\mu{}^m & = \tfrac12 \beps^i \Gamma^m \psi_{\mu i} \,,
  \notag \\
  \delta \psi_{\mu i} & = D_\mu \eps_i + \tfrac{1}{4 \sqrt{6}} i (
  \Gamma_\mu{}^{\nu \rho} - 4 \delta_{\mu}{}^\nu \Gamma^\rho) F_{\nu
  \rho} \eps_i \,, \notag \\
  \delta A_\mu & = - \tfrac{\sqrt{6}}{4} i \beps^i \psi_{\mu i} \,,
 \label{susy-transf-ung}
 \end{align}
where $i$ is the $SU(2)$ R-symmetry index.
These satisfy the commutator
 \begin{align}
  [\delta_1 , \delta_2] = \delta_{\rm gct} + \delta_{\rm Lorentz}
  + \delta_{\rm susy} + \delta_{\rm gauge}
  + \delta_{\mathcal L} \,,
 \label{susy-alg}
 \end{align}
where the first four terms on the right hand side are
transformations with the usual action on the fields and parameters
given by
 \begin{align}
  \xi^\mu & = \tfrac12 \beps_1^i \Gamma^\mu \eps_{2i} \,, \qquad
  \Lambda^{mn}  = \xi^\nu \omega_{\nu}{}^{mn} + \tfrac{1}{4
  \sqrt{6}} i \beps_1^i (\Gamma^{mnpq} + 4 g^{mp} g^{nq} ) F_{pq}
  \eps_{2i} \,, \notag \\
  \eta^i & = - \xi^\mu \psi_{\mu i} \,, \qquad
  \lambda^{(0)}  = - \tfrac{\sqrt{6}}{4} i \beps_1^i \eps_{2i} -
  \xi^\nu A_\nu \,,
 \end{align}
while $\delta_{\cal L}$ imposes a possible first-order field
equation (e.g.~for the gravitino) which takes the form of a duality relation
for the bosons.

In \cite{Gomis} it was shown that the supersymmetry algebra can also
be realised on higher-rank gauge potentials with the following
supersymmetry transformations:
 \begin{align}
  \delta B_{\mu \nu} & = b_1 \beps^i \Gamma_{[ \mu} \psi_{\nu ] i}
  + b_2 A_{[ \mu} \delta A_{\nu]} \,, \notag \\
  \delta C^{ij}_{\mu \nu \rho} & = i c_1 \beps^{(i} \Gamma_{[\mu \nu}
  \psi_{\rho]}^{j)} \,, \notag \\
  \delta D^{ij}_{\mu \nu \rho \sigma} & = d_1 \beps^{(i} \Gamma_{[ \mu \nu
  \rho} \psi_{\sigma]}^{j)} + d_2 A_{[\mu} \delta C^{ij}_{\nu \rho \sigma]} \,,
 \end{align}
provided $b_1 = \tfrac34 b_2 = - \tfrac12 \sqrt{6}$ and $c_1 d_2 = -
\sqrt{6} d_1$. The latter two are symmetric in their $SU(2)$ indices
and subject to the symplectic constraint
 \begin{align}
  C^{ij} - C_{ij}^* = D^{ij} - D_{ij}^* = 0 \,. \label{symplectic}
 \end{align}
They transform as $SU(2)$ triplets. On the right hand side of the
algebra we find the gauge transformations
 \begin{align} \label{transformation}
  \delta_{\rm gauge} B_{\mu \nu} & =
  - 2 \partial_{[ \mu} \lambda^{(1)}_{\nu ]} - \tfrac13 \sqrt{6} \lambda^{(0)} F_{\mu \nu}
  \,, \quad
  \lambda^{(1)}_\nu = - B_{\nu \sigma} \xi^\sigma + \tfrac14 \sqrt{6}
   \beps_1^i \Gamma_\nu \eps_{2i} - \tfrac12 i \beps_1^i \eps_{2i} A_\nu
   \,, \notag \\
  \delta_{\rm gauge} C^{ij}_{\mu \nu \rho} & =
  - 3 \partial_{[ \mu} \lambda^{(2)ij}_{\nu \rho ]} \,,
  \quad \lambda^{(2)ij}_{\mu \nu} = - C^{ij}_{\mu \nu \rho} \xi^\rho +
  \tfrac13 i c_1 \beps_1^{(i} \Gamma_{\mu \nu} \eps_2^{j)} \,, \\
  \delta_{\rm gauge} D^{ij}_{\mu \nu \rho \sigma} & =
  - 4 \partial_{[ \mu} \lambda^{(3)ij}_{\nu \rho \sigma]} \,,
   \quad \lambda^{(3)ij}_{\mu \nu \rho}  =
   - D^{ij}_{\mu \nu \rho \sigma} \xi^\sigma - \tfrac14 d_1 (\beps_1^{(i}
   \Gamma_{\mu \nu \rho} \eps_2^{j)} - \sqrt{6} i A_{[\mu} \beps_1^{(i}
   \Gamma_{\nu \rho ]} \eps_2^{j)}) \,. \notag
 \end{align}
and the following first-order field equations:
 \begin{align}
 \label{duality-relation}
   \delta_{\mathcal L} B_{\mu \nu} & = - (3 \partial_{[\mu} B_{\nu \rho]} - \sqrt{6}
   A_{[\mu} F_{\nu \rho]} - \tfrac{1}{2} \sqrt{-g}
   \varepsilon_{\mu \nu \rho
   \sigma \lambda} F^{\sigma \lambda}) \xi^\rho \,, \notag \\
  \delta_{\mathcal L} C^{ij}_{\mu \nu \rho} & = - (4 \partial_{[ \mu} C^{ij}_{\nu \rho \sigma]}) \xi^\sigma
  \,, \qquad
  \delta_{\mathcal L} D^{ij}_{\mu \nu \rho \sigma} =
  - (5 \partial_{[ \mu} D^{ij}_{\nu \rho \sigma \tau]}) \xi^\tau \,,
 \end{align}
The first line identifies $B$ as the dual of $A$ while the second
line implies the vanishing of the field strengths of $C^{ij}$ and
$D^{ij}$.

As shown in \cite{Gomis}, it is impossible to realise supersymmetry
on an independent five-form with a leading term that is bilinear in the gravitino and the supersymmetry parameter.
The only such possibility leads to the Levi-Civita tensor and hence a dependent field. However, it is in fact possible to introduce an $SU(2)$ triplet of five-forms whose supersymmetry variation contains only subleading terms that are trilinear in the gravitino, the supersymmetry parameter and a lower-rank gauge potential. Therefore
this supersymmetry variation will vanish in a linearised approximation. More precisely, their supersymmetry transformation is given by
 \begin{align}
  \delta E_{\mu \nu \rho \sigma \tau}^{ij} = e_3 B_{[\mu \nu} \delta
  C^{ij}_{\rho \sigma \tau]} + e_4 A_{[\mu} \delta D_{\nu \rho
  \sigma \tau]}^{ij} \,.
  \label{susy-5}
 \end{align}
The supersymmetry algebra closes on these five-forms provided $c_1
e_3 = \tfrac32 d_1 e_4$ and up to the gauge transformation
 \begin{align}
  & \delta_{\rm gauge} E_{\mu \nu \rho \sigma \tau}^{ij} = - 5
  \partial_{[\mu} \lambda^{(4)ij}_{\nu \rho \sigma \tau]} \,, \quad
  \lambda^{(4)ij}_{\mu \nu \rho \sigma} = - E^{(ij)}_{\mu \nu \rho \sigma \tau} \xi^\tau
  +  \tfrac15 d_1 e_4
  (\tfrac32 i B_{[\mu \nu} \bar{\epsilon}_1^{(i)} \Gamma_{\rho \sigma]} \epsilon_2^{j)} +
  A_{[\mu} \bar{\epsilon}_1^{(i)} \Gamma_{\nu \rho \sigma]} \epsilon_2^{j)} )\,.
 \end{align}
Note the absence of an independent four-form gauge transformation that is quadratic in the supersymmetry parameters and does not contain other gauge potentials. This follows from the unusual form of the supersymmetry transformations \eqref{susy-5}. It seems that such transformation properties have not been encountered before and are only possible for space-time filling top-forms.

\section{A semi-simple example in $\N=1$ supergravity} \label{sec:N=1}

\subsection{General aspects}

As an illustration of the general discussion in the conclusions
we will consider a particular semi-simple algebra
that appears naturally in $\N=1$ supergravity. As is well known, the
bosonic sector of the pure four-dimensional theory reduces to the (Ehlers)
scalar coset $SL(2) / SO(2)$ in three dimensions. Many aspects of
hidden symmetries were first discussed in this context. Instead of
the pure theory in four dimensions we consider the coupling to a
chiral multiplet, whose bosonic sector consists of two scalars that
can be taken to parametrise the scalar coset $SL(2) / SO(2)$ as
well. Consequently we will end up with the product of these two
simple factors in three dimensions.

\begin{table}[h!]
 \centering
\begin{tabular}{||c||c||c|c||}
\hline $D$ & $H_{\rm R}$ & $G$ & $H$ \\
\hline \hline
 4 & $SO(2)$ & $SO(n_{\rm V}) \times SL(2)$ & $SO(n_{\rm V}) \times SO(2)$ \\
\hline
 3 & $SO(2)$ & $SO(2,2+n_{\rm V})$ & $SO(2) \times SO(2+n_{\rm V})$  \\
\hline
    \end{tabular}
\caption{\sl The global symmetries $G$ and their compact subgroups $H$
of a class of $\N = 1$ theories, consisting in four dimensions of
the graviton, one chiral and $n_{\rm V}$ vector multiplets.}
\label{tab:N=1symm}
\end{table}

It will be useful to extend this theory further by adding $n_{\rm
V}$ vector multiplets with minimal couplings. The symmetries of
these theories and their dimensional reductions can be found in
table \ref{tab:N=1symm}. The theory without vectors, i.e.~$n_{\rm V}
= 0$, can be seen as a natural limit of the generic case with
$n_{\rm V} \geq 1$. However, note that that while $\fso(2,2+n_{\rm
V})$ is semi-simple for $n_{\rm V} =0$, it is simple for $n_{\rm V}
\geq 1$. Hence the very-extensions of the latter are well known. The
corresponding $(D-1)$- and $D$-form potentials\footnote{Note that
these are identical to the $(D-1)$- and $D$-form representations of
very-extended $SO(8,8+n_{\rm V})$, corresponding of half-maximal
supergravity \cite{half-maximal}, as these are different real forms
of the same algebra.} are given in table \ref{tab:N=1forms} in terms
of the global symmetry groups of these theories.

\begin{table}[ht]
 \centering
\begin{tabular}{||c||c|c||}
\hline $D$ & $(D-1)$-forms & $D$-forms \\
\hline \hline
  && \\[-10pt]
 $4$ & $(2,\fund) \oplus (2, \threeform)$ & $(3,1) \oplus (3, \twoform) \oplus (3, \fourform) \oplus \twoform \oplus \twoform \oplus \threehook$ \\[10pt]
\hline
  && \\[-10pt]
 $3$ & $1 \oplus \symm \oplus \fourform$ & $\symm \oplus \twoform \oplus \fourform \oplus \threehook \oplus \fivehook$ \\[13pt]
  \hline
    \end{tabular}
\caption{\sl The deformation and top--form potentials in terms of irreps of
  $G=SO(2,2+n_{\rm V})$ for this class of $\N = 1$
  supergravities. \label{tab:N=1forms}} 
\end{table}

The derivation of the representations in table \ref{tab:N=1forms} is
not valid for $n_{\rm V}=0$, since the corresponding Lie algebra is
semi-simple: $\fso(2,2) \simeq \fsl(2) \oplus \fsl(2)$. From the
supergravity viewpoint, however, one would of course expect a
`covariant' answer in $n_V$. Therefore one has the following consistency check
on any proposal for the very-extended $\fsl(2) \oplus \fsl(2)$: its
four- and three-dimensional decompositions should give rise to the
$(D-1)$- and $D$-forms of the above table in the limit where $n_{\rm
V}$ vanishes. We will now discuss this extended semi-simple Lie
algebra.

\subsection{Very-extended $\fsl(2) \oplus \fsl(2)$ and the $\N = 1$ theory}

\begin{figure}[h!]
\begin{center}
\begin{pspicture}(5,-1)(8,2)
\dualityNode{5,1}{N12042766640} \nodeLabel{N12042766640}{1}
\disabledNode{6,1}{N22042766640} \nodeLabel{N22042766640}{2}
\disabledNode{5,-1}{N32042766640} \nodeLabel{N32042766640}{3}
\normalNode{6,-1}{N42042766640} \nodeLabel{N42042766640}{4}
\normalNode{7,0}{N52042766640} \nodeLabel{N52042766640}{5}
\normalNode{8,0}{N62042766640} \nodeLabel{N62042766640}{6}
\singleConnection{N22042766640}{N52042766640}
\specialDoubleConnection{N22042766640}{N12042766640}
\singleConnection{N52042766640}{N62042766640}
\singleConnection{N52042766640}{N42042766640}
\specialDoubleConnection{N42042766640}{N32042766640}
\end{pspicture}
\end{center}
\caption{\sl $(\fsl(2) \oplus \fsl(2))^{+++}$ decomposed as
$\fsl(4)^{} \oplus \fsl(2)^{}$, corresponding to a $D=4$ theory with
an $SL(2)$ internal symmetry group.} \label{fig:A1A1-4D}
\end{figure}
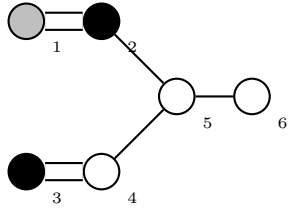

The decomposition of very-extended $\fsl(2) \oplus \fsl(2)$ with
respect to $\fsl(4)$, i.e.~in four dimensions, is illustrated in
figure \ref{fig:A1A1-4D}. The resulting list of generators with up
to four space-time indices is given in table \ref{fig:A1A1-4D} in
appendix \ref{sec:tables}. These can be interpreted to correspond to
the following physical degrees of freedom (at level $(l_1,l_2)$):
 \begin{itemize}
  \item
   $(0,0)$: the traceless part of the metric, carrying $15$ degrees of freedom
   of which $6$ will be eliminated due to the local Lorentz symmetry,
  \item
   $(0,0)$: the $\fsl(2)$ scalars, transforming in the adjoint of the internal
   symmetry and subject to the local $SO(2)$ symmetry,
  \item
   $(0,0)$: a scalar which provides the trace of the metric,
  \item
   $(1,0)$: a symmetric tensor, which is interpreted as the dual graviton,
  \item
   $(0,1)$: an $\fsl(2)$ triplet of two-forms, which are interpreted as dual to the scalars.
 \end{itemize}
Again one finds scalars and the dual two-forms in the adjoint of the
internal symmetry group $SL(2)$, while there are only corresponding
two physical degrees of freedom. This is taken care of by the
compact $SO(2)$ subgroup, which eliminates one scalar and imposes a
linear constraint on the three-form field strengths.

In this way the above generators exactly coincide with the physical
field content. Of the additional generators corresponding to
non-propagating degrees of freedom, the purely anti-symmetric ones
are
 \begin{itemize}
  \item
   $(0,2)$: an $\fsl(2)$ triplet of four-forms,
 \end{itemize}
while the very-extended algebra does not contain any three-forms.
These representations of three- and four-forms coincide with the
predictions of table \ref{tab:N=1forms} for $n_{\rm V} = 0$. As far
as we are aware, the possibility to include these four-forms in
$\N=1$, $D=4$ supergravity has not been discussed in the literature.

\begin{figure}[h!]
\begin{center}
\begin{pspicture}(5,-1)(8,2)
\dualityNode{5,1}{N12004488604} \nodeLabel{N12004488604}{1}
\disabledNode{6,1}{N22004488604} \nodeLabel{N22004488604}{2}
\dualityNode{5,-1}{N32004488604} \nodeLabel{N32004488604}{3}
\disabledNode{6,-1}{N42004488604} \nodeLabel{N42004488604}{4}
\normalNode{7,0}{N52004488604} \nodeLabel{N52004488604}{5}
\normalNode{8,0}{N62004488604} \nodeLabel{N62004488604}{6}
\singleConnection{N22004488604}{N52004488604}
\specialDoubleConnection{N22004488604}{N12004488604}
\singleConnection{N52004488604}{N62004488604}
\singleConnection{N52004488604}{N42004488604}
\specialDoubleConnection{N42004488604}{N32004488604}
\end{pspicture}
\end{center}
\caption{\sl $(\fsl(2) \oplus \fsl(2))^{+++}$ decomposed as
$\fsl(3)^{} \oplus \fsl(2) \oplus \fsl(2)$, corresponding to a $D=3$
theory with an $SO(2,2)$ internal symmetry group.}
\label{fig:A1A1-3D}
\end{figure}

Next we will discuss the decomposition of $(\fsl(2) \oplus
\fsl(2))^{+++}$ with respect to $\fsl(3)$, i.e.~in three dimensions.
The corresponding Dynkin diagram is given in figure
\ref{fig:A1A1-3D}, while the resulting list of generators can be
found in table \ref{fig:A1A1-3D} in appendix \ref{sec:tables}. It
can be verified that these lead to exactly the correct physical
degrees of freedom of the theory. Again the matching requires the
quotienting out of one of the central charges appearing at level
$(0,0)$ with multiplicity 2. In addition, the following
anti-symmetric non-propagating degrees of freedom are present in the
very-extended algebra:
 \begin{itemize}
  \item
  $l=(1,1)$, $(2,0)$ and $(0,2)$: two-forms in the $\fso(2,2)$ representations $\bf 1$ $\oplus$ $\bf 1$ $\oplus$ $\bf 9$,
  \item
   $l=(2,1)$ and $(1,2)$: three-forms in the $\fso(2,2)$ representations $\bf 3$ $\oplus$  $\bf 3$ $\oplus$  $\bf 9$ $\oplus$ $\bf 9$.
 \end{itemize}
Comparing with table \ref{tab:N=1forms}, the
representations of two-forms are again correct. The same holds for
the three-forms when a subtlety unrelated to the focus of this paper
has been taken into account\footnote{The general formulae would
predict an additional $\bf 1$. However, an additional requirement on
the three-forms is that their $\fso(2,2+n_{\rm V})$ representations
should be contained in the product of the representations of the
vectors and two-forms:
 \begin{align}
  \twoform \otimes (1 \oplus \symm \oplus \fourform) \,.
 \end{align}
For $n_{\rm V}$ large enough this contains the representations of
three-forms in table \ref{tab:N=1forms}. For $n_{\rm V}=0$ the
anti-symmetric four-form representation cannot be generated,
however. Hence the missing singlet in very-extended $\fsl(2) \oplus
\fsl(2)$ is not a problem: it should not be there, and indeed is
not. In fact, a missing top-form could be a more general phenomenon:
we are aware that there is a similar mismatch with respect to the
generic formulae for $\fso(4,4)^{+++}$ in $D=6$ and
$\fso(4,3)^{+++}$ in $D=5$. We thank Eric Bergshoeff and Teake Nutma
for discussions on this point.}.

\section{Decomposition tables} \label{sec:tables}

Below we give the relevant decompositions with respect to regular
$\fsl(D)_r$ subalgebras of the Kac--Moody algebras associated to very-extended semi-simple Lie algebras.
 The number of Cartan subalgebra scalars has been
adjusted according to the quotient procedure described in
section~\ref{sec:extensions}.  In all cases we
give all the generators at positive levels with up to and including
$D$ space-time indices. The tables were obtained by using the SimpLie program \cite{SimpLie} and Mathematica code by the first author which was also used for \cite{Kleinschmidt}.

The different columns in the tables are as follows:
\begin{itemize}
\item $l$ is the level in the decomposition,
\item $p_r$ are the Dynkin labels of the regular (gravity) subalgebra,
\item $p_i$ are the Dynkin labels of the internal symmetry subalgebra (if
  present),
\item `vector' is the root vector $\alpha$ in the algebra in whose root space the
  lowest weight vector of the representation lies,
\item $\alpha^2$ is the norm of the root vector,
\item $d_r$ is the dimension of the regular subalgebra representation,
\item $d_i$ is the dimension of the internal subalgebra representation,
\item $\mu$ is the outer multiplicity of the representation listed in a given
  row,
\item $ind$ is the number of space-time indices of this representation.
\end{itemize}

\begin{longtable}{|r@{\ }r|r@{\ }r@{\ }r@{\ }r|r@{\ }r@{\ }r@{\ }r|r@{\ }r@{\ }r@{\ }r@{\ }r@{\ }r@{\ }r@{\ }r@{\ }r@{\ }r|r|r|r|r|r|}
\hline \multicolumn{2}{|c|}{$l$} & \multicolumn{4}{|c|}{$p_r$} &
\multicolumn{4}{|c|}{$p_i$} & \multicolumn{10}{|c|}{vector $\alpha$} &
\multicolumn{1}{|c|}{$\alpha^2$} & \multicolumn{1}{|c|}{$d_r$} &
\multicolumn{1}{|c|}{$d_i$} & \multicolumn{1}{|c|}{$\mu$} &
\multicolumn{1}{|c|}{$ind$} \\
\hline \hline
0 & 0 & 0 & 0 & 0 & 0 & 0 & 0 & 0 & 1 & -2 & -4 & -3 & -2 & 0 & 0 & 0 & 0 & 0 & 0 & 12 & 1 & 52 & 1 & 0 \\
0 & 0 & 1 & 0 & 0 & 1 & 0 & 0 & 0 & 0 & 0 & 0 & 0 & 0 & 0 & 0 & -1 & -1 & -1 & -1 & 12 & 24 & 1 & 1 & 5  \\
0 & 0 & 0 & 0 & 0 & 0 & 0 & 0 & 0 & 0 & 0 & 0 & 0 & 0 & 0 & 0 & 0 & 0 & 0 & 0 & 0 & 1 & 1 & 1 & 0  \\
\hline
1 & 0 & 0 & 0 & 1 & 0 & 0 & 0 & 0 & 1 & 0 & 0 & 0 & 0 & 1 & 0 & 0 & 0 & 0 & 0 & 12 & 10 & 52 & 1 & 3  \\
\hline
0 & 1 & 1 & 0 & 0 & 0 & 0 & 0 & 0 & 0 & 0 & 0 & 0 & 0 & 0 & 1 & 0 & 0 & 0 & 0 & 4 & 5 & 1 & 1 & 1  \\
\hline
1 & 1 & 0 & 0 & 0 & 1 & 0 & 0 & 0 & 1 & 0 & 0 & 0 & 0 & 1 & 1 & 1 & 1 & 1 & 0 & 4 & 5 & 52 & 1 & 4 \\
\hline
0 & 2 & 0 & 1 & 0 & 0 & 0 & 0 & 0 & 0 & 0 & 0 & 0 & 0 & 0 & 2 & 1 & 0 & 0 & 0 & 4 & 10 & 1 & 1 & 2 \\
\hline
1 & 2 & 1 & 0 & 0 & 1 & 0 & 0 & 0 & 1 & 0 & 0 & 0 & 0 & 1 & 2 & 1 & 1 & 1 & 0 & 4 & 24 & 52 & 1 & 5 \\
1 & 2 & 0 & 0 & 0 & 0 & 0 & 0 & 0 & 1 & 0 & 0 & 0 & 0 & 1 & 2 & 2 & 2 & 2 & 1 & -8 & 1 & 52 & 1 & 5  \\
\hline
0 & 3 & 1 & 1 & 0 & 0 & 0 & 0 & 0 & 0 & 0 & 0 & 0 & 0 & 0 & 3 & 1 & 0 & 0 & 0 & 12 & 40 & 1 & 1 & 3  \\
\hline
0 & 4 & 1 & 0 & 1 & 0 & 0 & 0 & 0 & 0 & 0 & 0 & 0 & 0 & 0 & 4 & 2 & 1 & 0 & 0 & 4 & 45 & 1 & 1 & 4  \\
\hline
0 & 5 & 0 & 1 & 1 & 0 & 0 & 0 & 0 & 0 & 0 & 0 & 0 & 0 & 0 & 5 & 3 & 1 & 0 & 0 & 4 & 75 & 1 & 1 & 5  \\
0 & 5 & 1 & 0 & 0 & 1 & 0 & 0 & 0 & 0 & 0 & 0 & 0 & 0 & 0 & 5 & 3 & 2 & 1 & 0 & -8 & 24 & 1 & 1 & 5  \\
\hline
\caption{\label{tab:G2F4-5D}\sl $\fsl(5)_r^{} \oplus (\ff_{4})_{i}^{}$ representations in $(\fg_2 \oplus \ff_4)^{+++}$} 
\end{longtable}

\begin{longtable}{|r@{\ }r|r@{\ }r|r@{\ }r@{\ }r@{\ }r@{\ }r@{\ }r|r@{\ }r@{\ }r@{\ }r@{\ }r@{\ }r@{\ }r@{\ }r@{\ }r@{\ }r|r|r|r|r|r|}
\hline 
\multicolumn{2}{|c|}{$l$} & 
\multicolumn{2}{|c|}{$p_r$} & 
\multicolumn{6}{|c|}{$p_i$} & 
\multicolumn{10}{|c|}{vector $\alpha$} & 
\multicolumn{1}{|c|}{$\alpha^2$} & 
\multicolumn{1}{|c|}{$d_r$} & 
\multicolumn{1}{|c|}{$d_i$} & 
\multicolumn{1}{|c|}{$\mu$} & 
\multicolumn{1}{|c|}{$ind$}\\ 
\hline 
\hline 
0 & 0 & 0 & 0 & 0 & 0 & 0 & 0 & 0 & 0 & 0 & 0 & 0 & 0 & 0 & 0 & 0 & 0 & 0 & 0
& 0 & 1 & 1 & 1 & 0\\ 
0 & 0 & 1 & 1 & 0 & 0 & 0 & 0 & 0 & 0 & 0 & 0 & 0 & 0 & 0 & 0 & 0 & 0 & -1 &
-1 & 12 & 8 & 1 & 1 & 3\\  
0 & 0 & 0 & 0 & 0 & 0 & 0 & 0 & 0 & 1 & 0 & 0 & 0 & 0 & 0 & -3 & -2 & 0 & 0 &
0 & 12 & 1 & 14 & 1 & 0\\  
0 & 0 & 0 & 0 & 0 & 0 & 0 & 1 & 0 & 0 & -2 & -4 & -3 & -2 & 0 & 0 & 0 & 0 & 0
& 0 & 12 & 1 & 52 & 1 & 0\\  
\hline 
0 & 1 & 1 & 0 & 0 & 0 & 0 & 0 & 0 & 1 & 0 & 0 & 0 & 0 & 0 & 0 & 0 & 1 & 0 & 0
& 12 & 3 & 14 & 1 & 1\\  
\hline 
0 & 2 & 0 & 1 & 0 & 0 & 0 & 0 & 0 & 0 & 0 & 0 & 0 & 0 & 0 & 6 & 4 & 2 & 1 & 0
& -12 & 3 & 1 & 1 & 2\\ 
0 & 2 & 0 & 1 & 0 & 0 & 0 & 0 & 2 & 0 & 0 & 0 & 0 & 0 & 0 & 2 & 2 & 2 & 1 & 0
& 4 & 3 & 27 & 1 & 2\\ 
0 & 2 & 2 & 0 & 0 & 0 & 0 & 0 & 0 & 1 & 0 & 0 & 0 & 0 & 0 & 3 & 2 & 2 & 0 & 0
& 12 & 6 & 14 & 1 & 2\\ 
\hline 
1 & 0 & 1 & 0 & 0 & 0 & 0 & 1 & 0 & 0 & 0 & 0 & 0 & 0 & 1 & 0 & 0 & 0 & 0 & 0
& 12 & 3 & 52 & 1 & 1\\ 
\hline 
1 & 1 & 0 & 1 & 0 & 0 & 0 & 1 & 0 & 1 & 0 & 0 & 0 & 0 & 1 & 0 & 0 & 1 & 1 & 0
& 12 & 3 & 728 & 1 & 2\\ 
\hline 
2 & 0 & 0 & 1 & 0 & 0 & 0 & 0 & 0 & 0 & 4 & 8 & 6 & 4 & 2 & 0 & 0 & 0 & 1 & 0
& -12 & 3 & 1 & 1 & 2\\  
2 & 0 & 2 & 0 & 0 & 0 & 0 & 1 & 0 & 0 & 2 & 4 & 3 & 2 & 2 & 0 & 0 & 0 & 0 & 0
& 12 & 6 & 52 & 1 & 2\\  
2 & 0 & 0 & 1 & 2 & 0 & 0 & 0 & 0 & 0 & 0 & 2 & 2 & 2 & 2 & 0 & 0 & 0 & 1 & 0
& 12 & 3 & 324 & 1 & 2 \\ 
\hline 
1 & 2 & 0 & 0 & 0 & 0 & 0 & 1 & 0 & 0 & 0 & 0 & 0 & 0 & 1 & 6 & 4 & 2 & 2 & 1
& -24 & 1 & 52 & 1 & 3\\ 
1 & 2 & 1 & 1 & 0 & 0 & 0 & 1 & 0 & 0 & 0 & 0 & 0 & 0 & 1 & 6 & 4 & 2 & 1 & 0
& -12 & 8 & 52 & 1 & 3\\ 
1 & 2 & 0 & 0 & 0 & 0 & 0 & 1 & 0 & 1 & 0 & 0 & 0 & 0 & 1 & 3 & 2 & 2 & 2 & 1
& -12 & 1 & 728 & 1 & 3\\ 
1 & 2 & 0 & 0 & 0 & 0 & 0 & 1 & 2 & 0 & 0 & 0 & 0 & 0 & 1 & 2 & 2 & 2 & 2 & 1
& -8 & 1 & 1404 & 1 &3 \\ 
1 & 2 & 1 & 1 & 0 & 0 & 0 & 1 & 0 & 1 & 0 & 0 & 0 & 0 & 1 & 3 & 2 & 2 & 1 & 0
& 0 & 8 & 728 & 1 & 3\\ 
1 & 2 & 1 & 1 & 0 & 0 & 0 & 1 & 2 & 0 & 0 & 0 & 0 & 0 & 1 & 2 & 2 & 2 & 1 & 0
& 4 & 8 & 1404 & 1 & 3\\ 
1 & 2 & 0 & 0 & 0 & 0 & 0 & 1 & 3 & 0 & 0 & 0 & 0 & 0 & 1 & 0 & 1 & 2 & 2 & 1
& 12 & 1 & 4004 & 1 & 3\\ 
\hline 
2 & 1 & 0 & 0 & 0 & 0 & 0 & 0 & 0 & 1 & 4 & 8 & 6 & 4 & 2 & 0 & 0 & 1 & 2 & 1
& -24 & 1 & 14 & 1 & 3\\ 
2 & 1 & 1 & 1 & 0 & 0 & 0 & 0 & 0 & 1 & 4 & 8 & 6 & 4 & 2 & 0 & 0 & 1 & 1 & 0
& -12 & 8 & 14 & 1 & 3\\ 
2 & 1 & 0 & 0 & 0 & 0 & 0 & 1 & 0 & 1 & 2 & 4 & 3 & 2 & 2 & 0 & 0 & 1 & 2 & 1
& -12 & 1 & 728 & 1 & 3\\ 
2 & 1 & 1 & 1 & 0 & 0 & 0 & 1 & 0 & 1 & 2 & 4 & 3 & 2 & 2 & 0 & 0 & 1 & 1 & 0
& 0 & 8 & 728 & 1 & 3\\ 
2 & 1 & 0 & 0 & 2 & 0 & 0 & 0 & 0 & 1 & 0 & 2 & 2 & 2 & 2 & 0 & 0 & 1 & 2 & 1
& 0 & 1 & 4536 & 1 & 3\\ 
2 & 1 & 1 & 1 & 2 & 0 & 0 & 0 & 0 & 1 & 0 & 2 & 2 & 2 & 2 & 0 & 0 & 1 & 1 & 0
& 12 & 8 & 4536 & 1 & 3\\ 
2 & 1 & 0 & 0 & 0 & 0 & 1 & 0 & 0 & 1 & 0 & 0 & 0 & 1 & 2 & 0 & 0 & 1 & 2 & 1
& 12 & 1 & 17836 & 1 & 3\\ 
\hline 
\caption{\label{tab:G2F4-3D}\sl $\fsl(3)_r^{} \oplus (\ff_4 \oplus \fg_{2})_{i}^{}$ representations in $(\fg_2 \oplus \ff_4)^{+++}$} 
\end{longtable}

\begin{longtable}{|r@{\ }r|r@{\ }r@{\ }r@{\ }r@{\ }r|r@{\ }r@{\ }r|r@{\ }r@{\ }r@{\ }r@{\ }r@{\ }r@{\ }r@{\ }r@{\ }r@{\ }r|r|r|r|r|r|}
\hline \multicolumn{2}{|c|}{$l$} & \multicolumn{5}{|c|}{$p_r$} &
\multicolumn{3}{|c|}{$p_i$} & \multicolumn{10}{|c|}{vector $\alpha$} &
\multicolumn{1}{|c|}{$\alpha^2$} & \multicolumn{1}{|c|}{$d_r$} &
\multicolumn{1}{|c|}{$d_i$} & \multicolumn{1}{|c|}{$\mu$} &
\multicolumn{1}{|c|}{$ind$}\\
\hline \hline
0 & 0 & 1 & 0 & 0 & 0 & 1 & 0 & 0 & 0 & 0 & 0 & 0 & 0 & 0 & -1 & -1 & -1 & -1 & -1 & 12 & 35 & 1 & 1 & 6\\
0 & 0 & 0 & 0 & 0 & 0 & 0 & 0 & 1 & 0 & -3 & -2 & 0 & 0 & 0 & 0 & 0 & 0 & 0 & 0 & 12 & 1 & 14 & 1 & 0\\
0 & 0 & 0 & 0 & 0 & 0 & 0 & 0 & 0 & 2 & 0 & 0 & 0 & -1 & 0 & 0 & 0 & 0 & 0 & 0 & 6 & 1 & 3 & 1 & 0\\
0 & 0 & 0 & 0 & 0 & 0 & 0 & 0 & 0 & 0 & 0 & 0 & 0 & 0 & 0 & 0 & 0 & 0 & 0 & 0 & 0 & 1 & 1 & 1 & 0\\
\hline
1 & 0 & 0 & 0 & 0 & 1 & 0 & 0 & 1 & 0 & 0 & 0 & 1 & 0 & 0 & 0 & 0 & 0 & 0 & 0 & 12 & 15 & 14 & 1 & 4\\
\hline
0 & 1 & 1 & 0 & 0 & 0 & 0 & 0 & 0 & 1 & 0 & 0 & 0 & 0 & 1 & 0 & 0 & 0 & 0 & 0 & 6 & 6 & 2 & 1 & 1\\
\hline
1 & 1 & 0 & 0 & 0 & 0 & 1 & 0 & 1 & 1 & 0 & 0 & 1 & 0 & 1 & 1 & 1 & 1 & 1 & 0 & 6 & 6 & 28 & 1 & 5\\
\hline
0 & 2 & 0 & 1 & 0 & 0 & 0 & 0 & 0 & 2 & 0 & 0 & 0 & 0 & 2 & 1 & 0 & 0 & 0 & 0 & 12 & 15 & 3 & 1 & 2\\
\hline
1 & 2 & 1 & 0 & 0 & 0 & 1 & 0 & 1 & 2 & 0 & 0 & 1 & 0 & 2 & 1 & 1 & 1 & 1 & 0 & 12 & 35 & 42 & 1 & 6\\
1 & 2 & 0 & 0 & 0 & 0 & 0 & 0 & 1 & 2 & 0 & 0 & 1 & 0 & 2 & 2 & 2 & 2 & 2 & 1 & 0 & 1 & 42 & 1 & 6\\
1 & 2 & 0 & 0 & 0 & 0 & 0 & 0 & 1 & 0 & 0 & 0 & 1 & 1 & 2 & 2 & 2 & 2 & 2 & 1 & -6 & 1 & 14 & 1 & 6\\
\hline
0 & 3 & 0 & 0 & 1 & 0 & 0 & 0 & 0 & 1 & 0 & 0 & 0 & 1 & 3 & 2 & 1 & 0 & 0 & 0 & 6 & 20 & 2 & 1 & 3\\
\hline
0 & 4 & 1 & 0 & 1 & 0 & 0 & 0 & 0 & 0 & 0 & 0 & 0 & 2 & 4 & 2 & 1 & 0 & 0 & 0 & 12 & 105 & 1 & 1 & 4\\
0 & 4 & 0 & 0 & 0 & 1 & 0 & 0 & 0 & 2 & 0 & 0 & 0 & 1 & 4 & 3 & 2 & 1 & 0 & 0 & 6 & 15 & 3 & 1 & 4\\
\hline
0 & 5 & 1 & 0 & 0 & 1 & 0 & 0 & 0 & 1 & 0 & 0 & 0 & 2 & 5 & 3 & 2 & 1 & 0 & 0 & 6 & 84 & 2 & 1 & 5\\
0 & 5 & 0 & 0 & 0 & 0 & 1 & 0 & 0 & 3 & 0 & 0 & 0 & 1 & 5 & 4 & 3 & 2 & 1 & 0 & 6 & 6 & 4 & 1 & 5\\
0 & 5 & 0 & 0 & 0 & 0 & 1 & 0 & 0 & 1 & 0 & 0 & 0 & 2 & 5 & 4 & 3 & 2 & 1 & 0 & -6 & 6 & 2 & 1 & 5\\
\hline
0 & 6 & 0 & 1 & 0 & 1 & 0 & 0 & 0 & 2 & 0 & 0 & 0 & 2 & 6 & 4 & 2 & 1 & 0 & 0 & 12 & 189 & 3 & 1 & 6\\
0 & 6 & 1 & 0 & 0 & 0 & 1 & 0 & 0 & 2 & 0 & 0 & 0 & 2 & 6 & 4 & 3 & 2 & 1 & 0 & 0 & 35 & 3 & 1 & 6\\
0 & 6 & 1 & 0 & 0 & 0 & 1 & 0 & 0 & 0 & 0 & 0 & 0 & 3 & 6 & 4 & 3 & 2 & 1 & 0 & -6 & 35 & 1 & 2 & 6\\
0 & 6 & 0 & 0 & 0 & 0 & 0 & 0 & 0 & 4 & 0 & 0 & 0 & 1 & 6 & 5 & 4 & 3 & 2 & 1 & 6 & 1 & 5 & 1 & 0\\
0 & 6 & 0 & 0 & 0 & 0 & 0 & 0 & 0 & 2 & 0 & 0 & 0 & 2 & 6 & 5 & 4 & 3 & 2 & 1 & -12 & 1 & 3 & 2 & 0\\
\hline
\caption{\label{tab:G2F4-6D}\sl $\fsl(6)_r^{} \oplus (\fg_2 \oplus \fsl(2))_{i}^{}$ representations in $(\fg_2 \oplus \ff_4)^{+++}$} 
\end{longtable}

\newpage

\begin{longtable}{|r|r@{\ }r@{\ }r@{\ }r|r|r@{\ }r@{\ }r@{\ }r@{\ }r|r|r|r|r|r|r|}
\hline \multicolumn{1}{|c|}{$l$} & \multicolumn{4}{|c|}{$p_r$} &
\multicolumn{1}{|c|}{$p_i$} & \multicolumn{5}{|c|}{vector $\alpha$} &
\multicolumn{1}{|c|}{$\alpha^2$} & \multicolumn{1}{|c|}{$d_r$} &
\multicolumn{1}{|c|}{$d_i$} & \multicolumn{1}{|c|}{$\mu$} &
\multicolumn{1}{|c|}{$ind$} \\
\hline \hline
0 & 1 & 0 & 0 & 1 &  & 0 & -1 & -1 & -1 & -1 & 6 & 24 & 0 & 1 & 5  \\
0 & 0 & 0 & 0 & 0 &  & 0 & 0 & 0 & 0 & 0 & 0 & 1 & 0 & 1 & 0  \\
\hline
1 & 1 & 0 & 0 & 0 &  & 1 & 0 & 0 & 0 & 0 & 2 & 5 & 0 & 1 & 1  \\
\hline
2 & 0 & 1 & 0 & 0 &  & 2 & 1 & 0 & 0 & 0 & 2 & 10 & 0 & 1 & 2 \\
\hline
3 & 1 & 1 & 0 & 0 &  & 3 & 1 & 0 & 0 & 0 & 6 & 40 & 0 & 1 & 3  \\
\hline
4 & 1 & 0 & 1 & 0 &  & 4 & 2 & 1 & 0 & 0 & 2 & 45 & 0 & 1 & 4  \\
\hline
5 & 0 & 1 & 1 & 0 &  & 5 & 3 & 1 & 0 & 0 & 2 & 75 & 0 & 1 & 5  \\
5 & 1 & 0 & 0 & 1 &  & 5 & 3 & 2 & 1 & 0 & -4 & 24 & 0 & 1 & 5  \\
\hline
\caption{\label{tab:G2-5D} \sl $\fsl(5)_r^{}$ representations in $\fg_{2}^{+++}$} 
\end{longtable}

\begin{longtable}{|r@{\ }r|r@{\ }r@{\ }r@{\ }r|r|r@{\ }r@{\ }r@{\ }r@{\ }r@{\ }r@{\ }r|r|r|r|r|r|r|}
\hline \multicolumn{2}{|c|}{$l$} & \multicolumn{4}{|c|}{$p_r$} &
\multicolumn{1}{|c|}{$p_i$} & \multicolumn{7}{|c|}{vector $\alpha$} &
\multicolumn{1}{|c|}{$\alpha^2$} & \multicolumn{1}{|c|}{$d_r$} &
\multicolumn{1}{|c|}{$d_i$} & \multicolumn{1}{|c|}{$\mu$} &
\multicolumn{1}{|c|}{$ind$} \\
\hline \hline
0 & 0 & 1 & 0 & 0 & 1 & 0 & 0 & 0 & 0 & -1 & -1 & -1 & -1 & 6 & 24 & 1 & 1 & 5  \\
0 & 0 & 0 & 0 & 0 & 0 & 2 & -1 & 0 & 0 & 0 & 0 & 0 & 0 & 6 & 1 & 3 & 1 & 0  \\
0 & 0 & 0 & 0 & 0 & 0 & 0 & 0 & 0 & 0 & 0 & 0 & 0 & 0 & 0 & 1 & 1 & 1 & 0  \\
\hline
1 & 0 & 0 & 0 & 1 & 0 & 2 & 0 & 1 & 0 & 0 & 0 & 0 & 0 & 6 & 10 & 3 & 1 & 3  \\
\hline
0 & 1 & 1 & 0 & 0 & 0 & 0 & 0 & 0 & 1 & 0 & 0 & 0 & 0 & 2 & 5 & 1 & 1 & 1  \\
\hline
1 & 1 & 0 & 0 & 0 & 1 & 2 & 0 & 1 & 1 & 1 & 1 & 1 & 0 & 2 & 5 & 3 & 1 & 4  \\
\hline
0 & 2 & 0 & 1 & 0 & 0 & 0 & 0 & 0 & 2 & 1 & 0 & 0 & 0 & 2 & 10 & 1 & 1 & 2  \\
\hline
1 & 2 & 1 & 0 & 0 & 1 & 2 & 0 & 1 & 2 & 1 & 1 & 1 & 0 & 2 & 24 & 3 & 1 & 5  \\
1 & 2 & 0 & 0 & 0 & 0 & 2 & 0 & 1 & 2 & 2 & 2 & 2 & 1 & -4 & 1 & 3 & 1 & 5  \\
\hline
0 & 3 & 1 & 1 & 0 & 0 & 0 & 0 & 0 & 3 & 1 & 0 & 0 & 0 & 6 & 40 & 1 & 1 & 3  \\
\hline
0 & 4 & 1 & 0 & 1 & 0 & 0 & 0 & 0 & 4 & 2 & 1 & 0 & 0 & 2 & 45 & 1 & 1 & 4  \\
\hline
0 & 5 & 0 & 1 & 1 & 0 & 0 & 0 & 0 & 5 & 3 & 1 & 0 & 0 & 2 & 75 & 1 & 1 & 5  \\
0 & 5 & 1 & 0 & 0 & 1 & 0 & 0 & 0 & 5 & 3 & 2 & 1 & 0 & -4 & 24 & 1 & 1 & 5  \\
\hline
\caption{\label{tab:G2A1-5D} \sl $\fsl(5)_r^{} \oplus \fsu(2)_i^{}$ representations in $(\fg_2 \oplus \fsu(2))^{+++}$} 
\end{longtable}

\begin{longtable}{|r@{\ }r|r@{\ }r|r@{\ }r@{\ }r|r@{\ }r@{\ }r@{\ }r@{\ }r@{\ }r@{\ }r|r|r|r|r|r|r|}
\hline \multicolumn{2}{|c|}{$l$} & \multicolumn{2}{|c|}{$p_r$} &
\multicolumn{3}{|c|}{$p_i$} & \multicolumn{7}{|c|}{vector $\alpha$} &
\multicolumn{1}{|c|}{$\alpha^2$} & \multicolumn{1}{|c|}{$d_r$} &
\multicolumn{1}{|c|}{$d_i$} & \multicolumn{1}{|c|}{$\mu$} &
\multicolumn{1}{|c|}{$ind$} \\
\hline \hline
0 & 0 & 0 & 0 & 0 & 0 & 1 & 0 & 0 & -3 & -2 & 0 & 0 & 0 & 6 & 1 & 14 & 1 & 0  \\
0 & 0 & 1 & 1 & 0 & 0 & 0 & 0 & 0 & 0 & 0 & 0 & -1 & -1 & 6 & 8 & 1 & 1 & 3  \\
0 & 0 & 0 & 0 & 2 & 0 & 0 & -1 & 0 & 0 & 0 & 0 & 0 & 0 & 6 & 1 & 3 & 1 & 0  \\
0 & 0 & 0 & 0 & 0 & 0 & 0 & 0 & 0 & 0 & 0 & 0 & 0 & 0 & 0 & 1 & 1 & 2 & 0  \\
\hline
1 & 0 & 1 & 0 & 2 & 0 & 0 & 0 & 1 & 0 & 0 & 0 & 0 & 0 & 6 & 3 & 3 & 1 & 1  \\
\hline
0 & 1 & 1 & 0 & 0 & 0 & 1 & 0 & 0 & 0 & 0 & 1 & 0 & 0 & 6 & 3 & 14 & 1 & 1  \\
\hline
1 & 1 & 0 & 1 & 2 & 0 & 1 & 0 & 1 & 0 & 0 & 1 & 1 & 0 & 6 & 3 & 42 & 1 & 2  \\
\hline
2 & 0 & 2 & 0 & 2 & 0 & 0 & 1 & 2 & 0 & 0 & 0 & 0 & 0 & 6 & 6 & 3 & 1 & 2  \\
2 & 0 & 0 & 1 & 0 & 0 & 0 & 2 & 2 & 0 & 0 & 0 & 1 & 0 & -6 & 3 & 1 & 1 & 2  \\
\hline
0 & 2 & 2 & 0 & 0 & 0 & 1 & 0 & 0 & 3 & 2 & 2 & 0 & 0 & 6 & 6 & 14 & 1 & 2  \\
0 & 2 & 0 & 1 & 0 & 2 & 0 & 0 & 0 & 2 & 2 & 2 & 1 & 0 & 2 & 3 & 27 & 1 & 2  \\
0 & 2 & 0 & 1 & 0 & 0 & 0 & 0 & 0 & 6 & 4 & 2 & 1 & 0 & -6 & 3 & 1 & 1 & 2  \\
\hline
2 & 1 & 1 & 1 & 2 & 0 & 1 & 1 & 2 & 0 & 0 & 1 & 1 & 0 & 0 & 8 & 42 & 1 & 3  \\
2 & 1 & 1 & 1 & 0 & 0 & 1 & 2 & 2 & 0 & 0 & 1 & 1 & 0 & -6 & 8 & 14 & 1 & 3  \\
2 & 1 & 0 & 0 & 2 & 0 & 1 & 1 & 2 & 0 & 0 & 1 & 2 & 1 & -6 & 1 & 42 & 1 & 3  \\
2 & 1 & 0 & 0 & 0 & 0 & 1 & 2 & 2 & 0 & 0 & 1 & 2 & 1 & -12 & 1 & 14 & 1 & 3  \\
\hline
1 & 2 & 0 & 0 & 2 & 3 & 0 & 0 & 1 & 0 & 1 & 2 & 2 & 1 & 6 & 1 & 231 & 1 & 3  \\
1 & 2 & 1 & 1 & 2 & 2 & 0 & 0 & 1 & 2 & 2 & 2 & 1 & 0 & 2 & 8 & 81 & 1 & 3  \\
1 & 2 & 1 & 1 & 2 & 0 & 1 & 0 & 1 & 3 & 2 & 2 & 1 & 0 & 0 & 8 & 42 & 1 & 3  \\
1 & 2 & 0 & 0 & 2 & 2 & 0 & 0 & 1 & 2 & 2 & 2 & 2 & 1 & -4 & 1 & 81 & 1 & 3  \\
1 & 2 & 0 & 0 & 2 & 0 & 1 & 0 & 1 & 3 & 2 & 2 & 2 & 1 & -6 & 1 & 42 & 1 & 3  \\
1 & 2 & 1 & 1 & 2 & 0 & 0 & 0 & 1 & 6 & 4 & 2 & 1 & 0 & -6 & 8 & 3 & 1 & 3  \\
1 & 2 & 0 & 0 & 2 & 0 & 0 & 0 & 1 & 6 & 4 & 2 & 2 & 1 & -12 & 1 & 3 & 1 & 3  \\
\hline
3 & 0 & 3 & 0 & 2 & 0 & 0 & 2 & 3 & 0 & 0 & 0 & 0 & 0 & 6 & 10 & 3 & 1 & 3  \\
3 & 0 & 1 & 1 & 2 & 0 & 0 & 2 & 3 & 0 & 0 & 0 & 1 & 0 & -6 & 8 & 3 & 1 & 3  \\
3 & 0 & 1 & 1 & 0 & 0 & 0 & 3 & 3 & 0 & 0 & 0 & 1 & 0 & -12 & 8 & 1 & 1 & 3  \\
\hline
0 & 3 & 1 & 1 & 0 & 3 & 0 & 0 & 0 & 3 & 3 & 3 & 1 & 0 & 6 & 8 & 77 & 1 & 3  \\
0 & 3 & 3 & 0 & 0 & 0 & 1 & 0 & 0 & 6 & 4 & 3 & 0 & 0 & 6 & 10 & 14 & 1 & 3  \\
0 & 3 & 0 & 0 & 0 & 1 & 1 & 0 & 0 & 4 & 3 & 3 & 2 & 1 & -4 & 1 & 64 & 1 & 3  \\
0 & 3 & 1 & 1 & 0 & 2 & 0 & 0 & 0 & 5 & 4 & 3 & 1 & 0 & -4 & 8 & 27 & 1 & 3  \\
0 & 3 & 1 & 1 & 0 & 0 & 1 & 0 & 0 & 6 & 4 & 3 & 1 & 0 & -6 & 8 & 14 & 1 & 3  \\
0 & 3 & 0 & 0 & 0 & 2 & 0 & 0 & 0 & 5 & 4 & 3 & 2 & 1 & -10 & 1 & 27 & 1 & 3  \\
0 & 3 & 0 & 0 & 0 & 0 & 1 & 0 & 0 & 6 & 4 & 3 & 2 & 1 & -12 & 1 & 14 & 1 & 3  \\
0 & 3 & 1 & 1 & 0 & 1 & 0 & 0 & 0 & 7 & 5 & 3 & 1 & 0 & -10 & 8 & 7 & 1 & 3  \\
0 & 3 & 1 & 1 & 0 & 0 & 0 & 0 & 0 & 9 & 6 & 3 & 1 & 0 & -12 & 8 & 1 & 1 & 3  \\
\hline
\caption{\label{tab:G2A1-3D} \sl $\fsl(3)_r^{} \oplus (\fg_2 \oplus \fsu(2))_i^{}$ representations in $(\fg_2 \oplus \fsu(2))^{+++}$} 
\end{longtable}

\begin{longtable}{|r@{\ }r|r@{\ }r@{\ }r|r|r@{\ }r@{\ }r@{\ }r@{\ }r@{\ }r|r|r|r|r|r|}
\hline \multicolumn{2}{|c|}{$l$} & \multicolumn{3}{|c|}{$p_r$} &
\multicolumn{1}{|c|}{$p_i$} & \multicolumn{6}{|c|}{vector $\alpha$} &
\multicolumn{1}{|c|}{$\alpha^2$} & \multicolumn{1}{|c|}{$d_r$} &
\multicolumn{1}{|c|}{$d_i$} & \multicolumn{1}{|c|}{$\mu$} &
\multicolumn{1}{|c|}{$ind$} \\
\hline \hline
0 & 0 & 1 & 0 & 1 & 0 & 0 & 0 & 0 & -1 & -1 & -1 & 2 & 15 & 1 & 1 & 4  \\
0 & 0 & 0 & 0 & 0 & 2 & -1 & 0 & 0 & 0 & 0 & 0 & 2 & 1 & 3 & 1 & 0  \\
0 & 0 & 0 & 0 & 0 & 0 & 0 & 0 & 0 & 0 & 0 & 0 & 0 & 1 & 1 & 1 & 0  \\
\hline
1 & 0 & 0 & 1 & 0 & 2 & 0 & 1 & 0 & 0 & 0 & 0 & 2 & 6 & 3 & 1 & 2  \\
\hline
0 & 1 & 2 & 0 & 0 & 0 & 0 & 0 & 1 & 0 & 0 & 0 & 2 & 10 & 1 & 1 & 2  \\
\hline
1 & 1 & 1 & 0 & 1 & 2 & 0 & 1 & 1 & 1 & 1 & 0 & 0 & 15 & 3 & 1 & 4  \\
\hline
2 & 0 & 0 & 2 & 0 & 2 & 1 & 2 & 0 & 0 & 0 & 0 & 2 & 20 & 3 & 1 & 4  \\
2 & 0 & 1 & 0 & 1 & 0 & 2 & 2 & 0 & 0 & 1 & 0 & -2 & 15 & 1 & 1 & 4  \\
2 & 0 & 0 & 0 & 0 & 2 & 1 & 2 & 0 & 1 & 2 & 1 & -2 & 1 & 3 & 1 & 4  \\
\hline
0 & 2 & 2 & 1 & 0 & 0 & 0 & 0 & 2 & 1 & 0 & 0 & 2 & 45 & 1 & 1 & 4  \\
\hline
\caption{\label{tab:A1A1-4D}\sl $\fsl(4)_r^{} \oplus \fsl(2)_i^{}$ representations in $(\fsl(2) \oplus \fsl(2))^{+++}$} 
\end{longtable}

\newpage

\begin{longtable}{|r@{\ }r|r@{\ }r|r@{\ }r|r@{\ }r@{\ }r@{\ }r@{\ }r@{\ }r|r|r|r|r|r|}
\hline \multicolumn{2}{|c|}{$l$} & \multicolumn{2}{|c|}{$p_r$} &
\multicolumn{2}{|c|}{$p_i$} & \multicolumn{6}{|c|}{vector $\alpha$} &
\multicolumn{1}{|c|}{$\alpha^2$} & \multicolumn{1}{|c|}{$d_r$} &
\multicolumn{1}{|c|}{$d_i$} & \multicolumn{1}{|c|}{$\mu$} &
\multicolumn{1}{|c|}{$ind$} \\
\hline \hline
0 & 0 & 1 & 1 & 0 & 0 & 0 & 0 & 0 & 0 & -1 & -1 & 2 & 8 & 1 & 1 & 3  \\
0 & 0 & 0 & 0 & 0 & 2 & 0 & 0 & -1 & 0 & 0 & 0 & 2 & 1 & 3 & 1 & 0  \\
0 & 0 & 0 & 0 & 2 & 0 & -1 & 0 & 0 & 0 & 0 & 0 & 2 & 1 & 3 & 1 & 0  \\
0 & 0 & 0 & 0 & 0 & 0 & 0 & 0 & 0 & 0 & 0 & 0 & 0 & 1 & 1 & 1 & 0  \\
\hline
1 & 0 & 1 & 0 & 2 & 0 & 0 & 1 & 0 & 0 & 0 & 0 & 2 & 3 & 3 & 1 & 1  \\
\hline
0 & 1 & 1 & 0 & 0 & 2 & 0 & 0 & 0 & 1 & 0 & 0 & 2 & 3 & 3 & 1 & 1  \\
\hline
1 & 1 & 0 & 1 & 2 & 2 & 0 & 1 & 0 & 1 & 1 & 0 & 2 & 3 & 9 & 1 & 2  \\
\hline
2 & 0 & 2 & 0 & 2 & 0 & 1 & 2 & 0 & 0 & 0 & 0 & 2 & 6 & 3 & 1 & 2  \\
2 & 0 & 0 & 1 & 0 & 0 & 2 & 2 & 0 & 0 & 1 & 0 & -2 & 3 & 1 & 1 & 2  \\
\hline
0 & 2 & 2 & 0 & 0 & 2 & 0 & 0 & 1 & 2 & 0 & 0 & 2 & 6 & 3 & 1 & 2  \\
0 & 2 & 0 & 1 & 0 & 0 & 0 & 0 & 2 & 2 & 1 & 0 & -2 & 3 & 1 & 1 & 2  \\
\hline
2 & 1 & 1 & 1 & 2 & 2 & 1 & 2 & 0 & 1 & 1 & 0 & 0 & 8 & 9 & 1 & 3  \\
2 & 1 & 1 & 1 & 0 & 2 & 2 & 2 & 0 & 1 & 1 & 0 & -2 & 8 & 3 & 1 & 3  \\
2 & 1 & 0 & 0 & 2 & 2 & 1 & 2 & 0 & 1 & 2 & 1 & -2 & 1 & 9 & 1 & 3  \\
2 & 1 & 0 & 0 & 0 & 2 & 2 & 2 & 0 & 1 & 2 & 1 & -4 & 1 & 3 & 1 & 3  \\
\hline
1 & 2 & 1 & 1 & 2 & 2 & 0 & 1 & 1 & 2 & 1 & 0 & 0 & 8 & 9 & 1 & 3  \\
1 & 2 & 1 & 1 & 2 & 0 & 0 & 1 & 2 & 2 & 1 & 0 & -2 & 8 & 3 & 1 & 3  \\
1 & 2 & 0 & 0 & 2 & 2 & 0 & 1 & 1 & 2 & 2 & 1 & -2 & 1 & 9 & 1 & 3  \\
1 & 2 & 0 & 0 & 2 & 0 & 0 & 1 & 2 & 2 & 2 & 1 & -4 & 1 & 3 & 1 & 3  \\
\hline
3 & 0 & 3 & 0 & 2 & 0 & 2 & 3 & 0 & 0 & 0 & 0 & 2 & 10 & 3 & 1 & 3  \\
3 & 0 & 1 & 1 & 2 & 0 & 2 & 3 & 0 & 0 & 1 & 0 & -2 & 8 & 3 & 1 & 3  \\
3 & 0 & 1 & 1 & 0 & 0 & 3 & 3 & 0 & 0 & 1 & 0 & -4 & 8 & 1 & 1 & 3  \\
\hline
0 & 3 & 3 & 0 & 0 & 2 & 0 & 0 & 2 & 3 & 0 & 0 & 2 & 10 & 3 & 1 & 3  \\
0 & 3 & 1 & 1 & 0 & 2 & 0 & 0 & 2 & 3 & 1 & 0 & -2 & 8 & 3 & 1 & 3  \\
0 & 3 & 1 & 1 & 0 & 0 & 0 & 0 & 3 & 3 & 1 & 0 & -4 & 8 & 1 & 1 & 3  \\
\hline
\caption{\label{tab:A1A1-3D}\sl $\fsl(3)_r^{} \oplus \fsl(2)_i^{} \oplus \fsl(2)_i^{}$ representations in $(\fsl(2) \oplus \fsl(2))^{+++}$} 
\end{longtable}

  \end{document}